\documentclass[useAMS,usenatbib]{mn2e}

\pdfoutput=1

\usepackage{graphicx, fixltx2e, subfigure, color, placeins, amsmath, url}

\graphicspath{ {/Users/salberts/Research/Data/Bootes/Herschel/SPIRE/Stacking/Weighted/} }
\providecommand{\e}[1]{\ensuremath{\times 10^{#1}}}
\newcommand{\Msun}{\ensuremath{\mathrm{M_{\odot}}}}
\newcommand{\Lsun}{\ensuremath{\mathrm{L_{\odot}}}}
\newcommand{\avg}[1]{\langle #1 \rangle}
\newcommand {\lesssim}
  {{\hbox{\,\rlap{\hbox{\lower4pt\hbox{$\sim$}}}\hbox{$<$}}\,}}
\newcommand {\gtrsim}
  {{\hbox{\,\rlap{\hbox{\lower4pt\hbox{$\sim$}}}\hbox{$>$}}\,}}

\title[Star formation in galaxy clusters]{\vspace{-0.2in}The evolution of dust-obscured star formation activity in galaxy clusters relative to the field over the last 9 billion years\thanks{{\it Herschel} is an ESA space observatory with science instruments provided by European-led Principal Investigator consortia and with important participation from NASA.}\vspace{-0.2in}}
\author[Alberts et al.]{\parbox{\textwidth}{
Stacey Alberts$^1$,
Alexandra Pope$^1$, 
Mark Brodwin$^2$, 
David W. Atlee$^3$, 
Yen-Ting Lin$^4$,
Arjun Dey$^3$, 
Peter R. M. Eisenhardt$^5$, 
Daniel P. Gettings$^6$, 
Anthony H. Gonzalez$^6$,
Buell T. Jannuzi$^7$,
Conor L. Mancone$^6$,
John Moustakas$^8$,
Gregory F. Snyder$^9$,   
S. Adam Stanford$^{10}$, 
Daniel Stern$^5$,
Benjamin J. Weiner$^{7}$,
Gregory R. Zeimann$^{11}$  }\vspace{0.2cm}\\
$^1$Department of Astronomy, University of Massachusetts, LGRT-B 619E, Amherst, MA 01003, USA \\
$^2$Department of Physics and Astronomy, University of Missouri, 5110 Rockhill Road, Kansas City, MO 64110 \\
$^3$National Optical Astronomy Observatory, 950 N. Cherry Ave., Tucson, AZ 85719 \\
$^4$Institute of Astronomy and Astrophysics, Academia Sinica, Taipei, Taiwan \\
$^5$Jet Propulsion Laboratory, California Institute of Technology, Pasadena, CA 91109 \\
$^6$ Department of Astronomy, University of Florida, Gainesville, FL 32611-2055 \\
$^7$ Steward Observatory, 933 N. Cherry St., University of Arizona, Tuscon, AZ 85721, USA \\
$^8$Department of Physics \& Astronomy, Siena College, 515 Loudon Road, Loudonville, NY 12211 \\
$^9$  Havard-Smithsonian Center for Astrophysics, 60 Garden St. Cambridge, MA 02138 \\
$^{10}$University of California, Davis, CA 95616 \\
$^{11}$Department of Astronomy and Astrophysics, Pennsylvania State University, 525 Davey Laboratory, University Park, Pennsylvania 16802  \vspace{-0.2in}}
\date{Updated \today}
\pagerange{\pageref{firstpage}--\pageref{lastpage}} \pubyear{2013}
\begin{document}
\label{firstpage}
\maketitle
\begin{abstract}
We compare the star formation (SF) activity in cluster galaxies to the field from $z=0.3-1.5$ using $Herschel$ Spectral and Photometric Imaging REceiver (SPIRE) 250$\mu$m imaging.  We utilize 274 clusters from the IRAC Shallow Cluster Survey (ISCS) selected as rest-frame near-infrared overdensities over the 9 square degree Bo\"{o}tes field .   This analysis allows us to quantify the evolution of SF in clusters over a long redshift baseline without bias against active cluster systems.  Using a stacking analysis, we determine the average star formation rates (SFRs) and specific-SFRs (SSFR=SFR/M$_{\star}$) of stellar mass-limited (M$\,\geq1.3\e{10}\Msun$), statistical samples of cluster and field galaxies, probing both the star forming and quiescent populations. We find a clear indication that the average SF in cluster galaxies is evolving more rapidly than in the field, with field SF levels at $z\gtrsim1.2$ in the cluster cores ($r<0.5\,$Mpc), in good agreement with previous ISCS studies.  By quantifying the SF in cluster and field galaxies as an exponential function of cosmic time, we determine that cluster galaxies are evolving $\sim2$ times faster than the field.  Additionally, we see enhanced SF above the field level at $z\sim1.4$ in the cluster outskirts ($r>0.5\,$Mpc).  These general trends in the cluster cores and outskirts are driven by the lower mass galaxies in our sample. Blue cluster galaxies have systematically lower SSFRs than blue field galaxies, but otherwise show no strong differential evolution with respect to the field over our redshift range.  This suggests that the cluster environment is both suppressing the star formation in blue galaxies on long time-scales and rapidly transitioning some fraction of blue galaxies to the quiescent galaxy population on short time-scales.  We argue that our results are consistent with both strangulation and ram pressure stripping acting in these clusters, with merger activity occurring in the cluster outskirts.
\end{abstract}


\FloatBarrier

\newpage
\section{Introduction}
\label{sec:intro}

It is well established that in the local Universe galaxy properties are strongly correlated with both their local environment and their stellar mass \citep[e.g.,][]{pen10}.  Local clusters host strong red sequences of passively evolving galaxies with little to no star formation (SF),  while the lower density field contains the bulk of star forming galaxies \citep[see][for a review]{bla09}.  Similarly, massive galaxies tend to be redder, with old galaxy populations and low star formation rates  \citep[SFRs; e.g.][]{bow92, bal06, wei06, pen10, tho10}.  Massive galaxies are also known to reside preferentially in denser environments \citep{kau04, bal06}.  So while it is clear that environment plays a prominent role in galaxy evolution, it is still controversial whether the role of environment is direct, operating through processes external to individual galaxies and specific to dense regions, or indirect, with galaxy density tracing specific galaxy populations (such as massive galaxies) whose evolution is dominated by their own internal mechanisms.   Given that environmental effects are also likely strongly dependent on cosmic time in an evolving Universe,  it is important to quantify the transition epoch from active star formation and mass assembly to passive evolution in the densest environments.  
%, with higher redshift cluster studies revealing a transition to active cluster assembly at $z\sim1.5$ \citep{man10, bro13}. 

Cluster studies have determined that the density local correlations are in place at $z\sim1$ \citep[e.g.,][]{pat09, muz12}.  Recently, \citet{sco13} analysed a large dynamical range of environments in the Cosmic Evolution Survey (COSMOS) field and determined that the strong correlation between red, passive galaxies and dense environments becomes much weaker at $z>1.2$.  Though \citet{sco13} and other studies \citep{pat09, cuc10, bol10} did not observe a reversal of the local SFR-density relation (where SF decreases with increasing galaxy density up to group scales) as found previously \citep[][see also Cooper et al. 2008]{elb07}, multiple high redshift studies of galaxy clusters have presented tantalizing evidence of increased star formation activity toward the densest regions.  Infrared (IR) studies have noted increasing fractions of Luminous Infrared Galaxies (LIRGs; $1\e{11}\,\Lsun<{\rm L_{\rm IR}}<1\e{12}\,\Lsun$) and Ultra-Luminous Infrared Galaxies (ULIRGs; L$_{\rm IR}>1\e{12}\,\Lsun$)  in clusters out to $z\sim0.8$ \citep{coi05, gea06, mar07, muz08, koy08, hai09, smi10, chu11}.  Studies of the evolution of cluster galaxies up to $z\sim1$ have found increasing fractions of star forming galaxies in cluster cores \citep{sai08, web13, hai13} and the total SFR per unit halo mass in clusters has been found to be evolving as fast or faster than the field with a redshift dependence of roughly (1$+z)^{5-7}$  \citep{kod04, bai09, pop12, web13, hai13}.   At higher redshifts, individual cluster studies have revealed increased star formation activity down into the cluster cores \citep[$z>1.4$;][]{tra10, hil10, hay11, fas11, tad12}.   Small cluster samples, however, are susceptible to large variations in clusters properties \citep{gea06} and these works highlighted the need for evolutionary studies of large, uniform cluster samples over a long redshift baseline.   Recently, such studies have shown active mass assembly in clusters  \citep{man10}, stochastic star formation histories \citep{sny12}, and a transition to active star formation in clusters at high redshift \citep{bro13}.

The mechanisms which drive the majority of cluster galaxies from actively star forming to passively evolving have not yet been fully identified.  Multiple interpretations have been put forth as to the environment's role in the suppression of star formation.  \citet{pen10} found that the effects of environment and the stellar mass of galaxies are largely separable at $z\sim1$, with the environment playing no substantial role in the quenching process for massive galaxies, whose evolution is dominated by internal self-quenching (so-called mass-quenching).  \citet{muz12} found that the specific star formation rates (SSFR=SFR/M$_{\star}$) of star forming galaxies appear independent of environment and interpreted  the environment's primary function as controlling the fraction of star forming to quiescent galaxies through quenching on rapid time-scales.  This is further supported by differences found in the stellar mass distributions of cluster and field galaxy populations \citep{van13}.  Studies of the 3.6 and 4.5$\mu$m luminosity function in clusters found evidence for mass assembly at high redshift \citep{man10}, which is consistent with the two order of magnitude increase in active galactic nucleus (AGN) activity in cluster galaxies from $z=0-1.5$ \citep{gal09, mar13} and may indicate a prominent role for mergers in cluster environments.  More long redshift baseline studies of large, uniform cluster catalogues are necessary to quantify the relative importance of mass- versus environmental-quenching as well as what cluster-specific processes may drive the evolution of cluster populations.

In addition to needing large cluster samples over a range of redshifts, studies have shown that the prominence  of dust-obscured star formation increases with redshift, with the majority of star formation enshrouded by dust at $z>1$ \citep[e.g.,][]{bou09, mag13}.  Infrared observations of clusters are therefore necessary to get a complete census of star formation over a large redshift range.  Current mid-IR studies of clusters \citep[e.g.][]{web13, bro13} have analysed detected infrared sources and have thus probed relatively bright IR galaxy populations.  Complimentary to this, a stacking analysis can measure average star formation properties by probing farther down the luminosity function, including relatively quiescent galaxies, for a look at the full population of cluster galaxies. 

In this study, we quantify the average star formation properties of cluster galaxies over a long baseline of cosmic time out to $z=1.5$ ($\sim9$ billion years ago) using a uniform, stellar mass-selected sample of 274 clusters over the 9 square degree Bo\"{o}tes field.  This is the first study to measure the star formation properties in stellar mass-limited cluster and field galaxy samples over such a long redshift baseline.  Our cluster sample is identified as three-dimensional near-infrared overdensities in photometric redshift space; as such, we do not rely on the presence of absence of a red sequence and thus are not biased against actively forming clusters.   Cluster membership is determined using spectroscopic and photometric redshifts and we perform a robust, statistical removal of contaminating field galaxies.  The cluster SF properties are compared to those of a field galaxy sample drawn from the same 4.5$\mu$m-selected catalogue.  Stellar masses are available for our entire catalogue enabling us to construct stellar mass-limited galaxy samples.  SFRs and SSFRs are obtained by a stacking analysis performed on $Herschel$ SPIRE 250$\mu$m imaging.  By stacking thousands of cluster galaxies and tens of thousands of field galaxies, we derive robust measurements of the average 250$\mu$m flux, from which we derive accurate estimates of the L$_{\rm IR}$ and dust-obscured SFR.    Our stacking analysis accounts for the contribution from both star forming and quiescent galaxies.  Given our large samples of cluster and field galaxies, we are able to break our analysis down into subsets by stellar mass and galaxy colour.    By quantifying the rate of evolution out to high redshift, we constrain which processes might dominate the change in cluster galaxy properties and present arguments for specific quenching mechanisms in the clusters.

In Section 2, we present our cluster sample, cluster and field membership selection, and describe the $Herschel$ SPIRE imaging and other ancillary data used.  In Section 3, we lay out the stacking analysis, including the stacking procedure at 250$\mu$m and our method for stacking clusters members including corrections for source blending/clustering bias and field contamination.  We discuss the procedure for stacking field galaxies, and our report on possible complications from projection effects and AGN.  This section also includes the procedure for stacking at 70$\mu$m, a check on possible systematics introduced during the conversion of 250$\mu$m flux to L$_{\rm IR}$.  In Section 4, we detail the conversion of 250$\mu$m fluxes to galaxy properties (L$_{\rm IR}$ and SFR) and present the results of the stacking analysis for cluster and field galaxy samples.  In Section 5, we discuss our results in terms of environmental and internal quenching mechanisms, place our results in the context of other studies.  Section 6 contains our conclusions.  Throughout this work,  we adopt a Wilkinson Microwave Anisotropy Probe (WMAP) 7 cosmology with ($\Omega_{\Lambda}, \Omega_{\textup{M}}, h$)=(0.728, 0.272, 0.704) \citep{kom11}.


\section{Data}
\label{sec:data}
\subsection{ISCS Cluster Sample}
\label{sec:clusters}

The IRAC Shallow Cluster Survey \citep[ISCS;][]{eis08} is a sample of 335 clusters over the redshift range $0<z<2$ (106 at $z>1$) in the Bo\"{o}tes field.  Clusters were identified via a wavelet search algorithm which determined statistically significant rest-frame near-infrared overdensities in three-dimensional redshift slices using the photometric redshift probability distribution functions of 4.5$\mu$m-selected galaxies across the field.  The photometric redshifts used for cluster identification \citep{bro06} were calculated using deep $B_W$, $R$, and $I$ band optical data from the NOAO Deep, Wide-Field Survey \citep[NDWFS;][]{jan99} and $Spitzer$ IRAC 3.6 and 4.5$\mu$m imaging from the IRAC Shallow Survey \citep[ISS;][]{eis04}.  As the ISS is 4.5$\mu$m flux-limited (8.8$\mu$Jy at 5$\sigma$), this cluster sample is essentially stellar mass selected and does not require nor preclude the presence of a strong red sequence in the clusters.   Spectroscopic confirmation of dozens of the ISCS clusters at low redshifts ($z\leq0.9$) was obtained through the AGN and Galaxy Evolution Survey \citep[AGES;][]{koc12} and the Low Resolution Imaging Spectrometer (LRIS) on Keck  \citep{ste10}.  Additionally, over 20 of the clusters at $z>$1 have been spectroscopically confirmed via Keck or $Hubble$ $Space$ $Telescope$ spectroscopy \citep{sta05, sta12, els06, bro06, bro11, eis08, zei12, bro13, zei13}.  Overall, this cluster sample is expected to have a $\sim10\%$ false detection rate due to chance projections \citep{eis08}.

In order to characterize the ISCS cluster masses as a function of redshift, we perform a halo mass ranking simulation following the procedure in \citet{lin13}.  We determine the median mass of the $N$ most luminous clusters, as determined from their total 4.5$\mu$m luminosity, in redshift bins with width 0.2 from z=0.3-1.5 (see Table~\ref{tbl:stats}).  We find a range of median cluster masses of M$_{200}\sim5\e{13}-8\e{13}\,\Msun$ with no significant evolution with redshift.  This is consistent with measurements of individual ISCS clusters of the dynamical \citep{sta05, els06, bro06, eis08} and X-ray \citep{bro11} masses, as well as an analysis of the galaxy cluster autocorrelation function for the ISCS sample \citep{bro07} which found the characteristic cluster mass to be $\sim10^{14}\,\Msun$.   Mass estimates from weak lensing are also available for six ISCS clusters at $z>1$  \citep{jee11}.

In this study, we analyse the star formation properties of the 274 clusters that fall within the coverage of the SPIRE Bo\"{o}tes maps (Section~\ref{sec:fir}), presenting the evolution of a uniform cluster sample with cosmic time and redshift.  These clusters span the redshift range $0.3<z<1.5$, over which the photometric redshifts have a uniform accuracy as described in the next section.

%An analysis of the galaxy cluster autocorrelation function of the ISCS clusters indicates a typical halo mass of $\sim10^{14}$ $\Msun$ across the sample \citep{bro07}.  This has been further confirmed by measurements of the dynamical \citep{sta05, els06, bro06, eis08}, X-ray \citep{bro11}, and weak lensing \citep{jee11} masses of a subset of the clusters.  In this study, we analyse the star formation properties of the 274 clusters which fall within the field of view (FOV) of the SPIRE B\"{o}otes maps (Section~ref{sec:fir}).  This sample spans the redshift range 0.3$<z<$1.5, over which the photometric redshifts have a uniform accuracy as described in the next section.

\subsubsection{The Bo\"{o}tes Field:  photometric Redshifts and Stellar Mass Estimates}
\label{sec:redshifts}

Photometric redshifts are available across the 9 square degree NDWFS Bo\"{o}tes field, which includes the SPIRE Bo\"{o}tes imaging.  The photometric redshifts used in this study were updated from the original work in \citet{bro06} to incorporate infrared data from the Spitzer Deep, Wide-Field Survey \citep[SDWFS;][]{ash09}, which repeated the 90 second exposure of the ISS three more times.  This resulted in a factor of two increase in the catalogue depth, a significantly more robust catalogue with regards to cosmic rays and instrumental effects, and a greater sensitivity to distant galaxies in the 5.8 and 8.0$\mu$m bands.  Photometric redshifts for 434,295 galaxies were determined by fitting a subset of models (late types: Sb, Sc, Sd, Spi4, M82; early types: Ell5, Ell13, S0 and Sa) from \citet{pol07} to rest-frame wavelengths $\sim0.1-8\mu$m over $0<z<2$.  These models were chosen over the original models used in \citet{bro06} as they span the full wavelength range probed by the NDWFS+SDWFS filters \citep[see][for more details]{bro13}. A comparison with available spectroscopic redshifts shows that the precision of these photometric redshifts is $\sigma$ $\sim0.06(1+z$) for 95$\%$ of the galaxies over the redshift range $0<z<1.5$ \citep{bro13}.  To be conservative, we further limit our lower redshift bound to $z\geq0.3$, below which the 4000\AA break is blueward of the NDWFS filters.

Estimates of the stellar masses for galaxies in the photometric redshift catalogue were calculated using {\tt iSEDfit} \citep{mou13}, a Bayesian spectral energy distribution (SED )fitting code.  The data are fit using the \citet{bru03} population synthesis models and assuming the \citet{cha03} initial mass function (IMF) from 0.1-100 $\Msun$.  We adopt a stellar mass cutoff of $M_{\star} = 1.3\e{10} \Msun$ throughout this work, which corresponds to the mass limit of our sample at $z$=1.5 \citep[see Figure~3 in][]{bro13}.  Though the statistical uncertainties in the stellar masses are typically $\lesssim0.2$ dex, we adopt a conservative error of 0.3 dex on all stellar masses to account for systematic uncertainties \citep[see Appendix A][for a more in-depth discussion of how our stellar masses are derived]{mou13}.

\subsubsection{Cluster Membership}
\label{sec:membership}

Cluster membership is first determined through available spectroscopic redshifts.  As in \citet{eis08}, if a spectroscopic redshift is within 2000 km s$^{-1}$ of the systemic cluster velocity and lies within a 2 Mpc radius of the projected cluster center, it is considered to be a cluster member.  The cluster centers are taken from the density peaks identified by the wavelet search algorithm.  

\begin{figure}
\hspace{-5mm}
\includegraphics[scale=0.09, trim=18mm 20mm 20mm 25mm, clip]{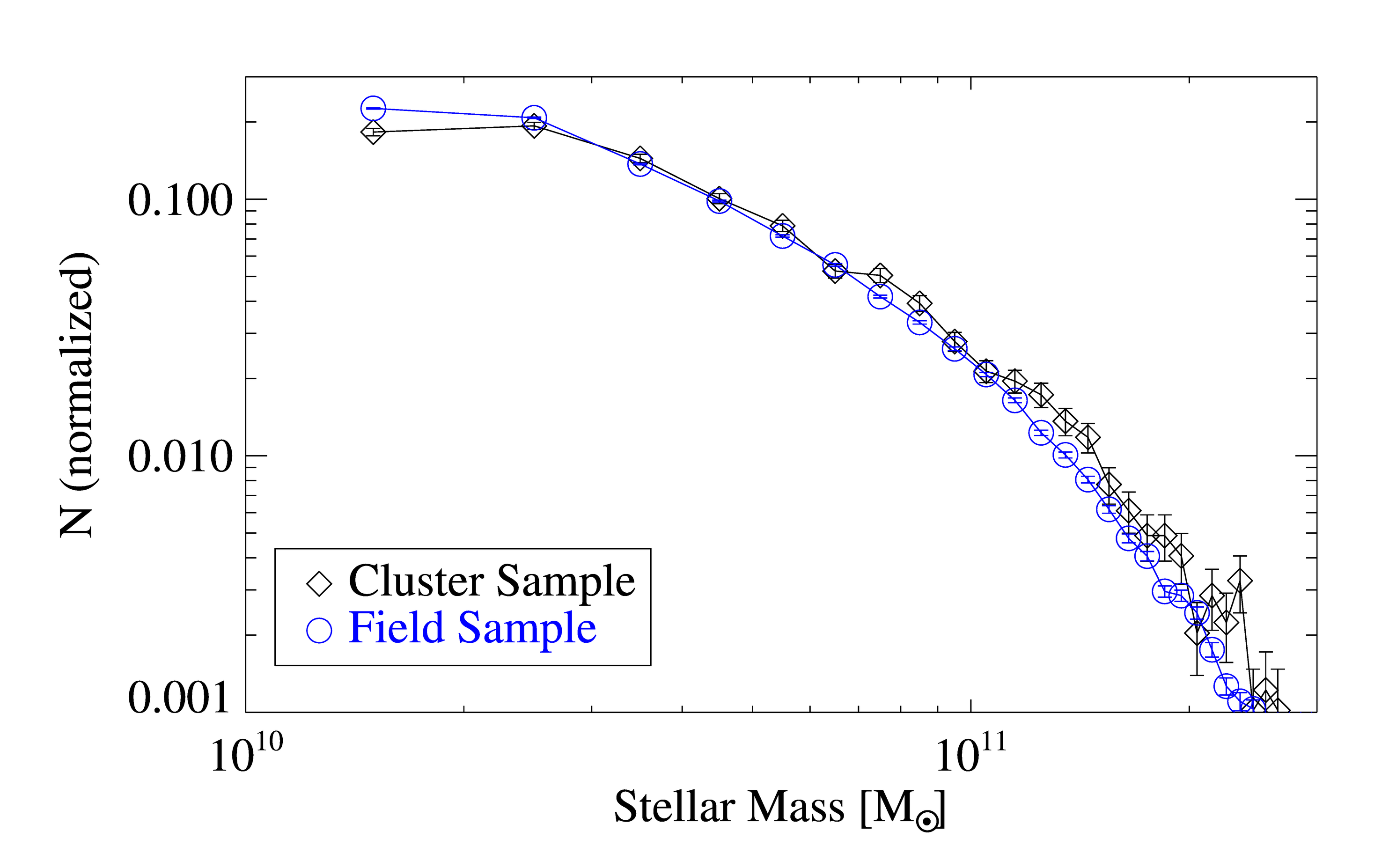}
\caption{The stellar mass distribution of our cluster (black diamonds) and field (blue circles) galaxy samples as described in Sections~\ref{sec:membership} and ~\ref{sec:field}.  Each distribution includes galaxies from $z=0.3-1.5$ and is normalized by the total number of galaxies in each sample.  The cluster sample has not been corrected for field contamination (see Section~\ref{sec:clusterstacks}).  Given the uncertainties, the stellar mass distributions of cluster and field galaxies are similar over most masses, with slightly fewer cluster galaxies in the lowest mass bin and correspondingly more cluster galaxies at higher masses.  We see these same small differences when we split our samples at $z=1$, though we do display these plots due to poor statistics.}
\label{fig:massdist}
\end{figure}

Galaxies with only photometric redshifts are assigned membership based on a constraint of the integral of their normalized photometric redshift probability distributions:

\begin{equation}
\label{eqn:membership}
\int_{z_{cl}-0.06(1+z_{cl})}^{z_{cl}+0.06(1+z_{cl})} P(z) dz \geq 0.3
\end{equation}
where $z_{cl}$ is the redshift of the cluster, calculated by iteratively summing up the $P(z)$ function for potential cluster members within 1~Mpc and re-identifying cluster members until convergence.  Galaxies which satisfy Eqn~\ref{eqn:membership} and are within 2 Mpc of the projected cluster center are photometric redshift cluster members.  The numbers of spectroscopic and photometric cluster members for $r\leq1\,$Mpc (approximately the virial radius) can be seen in Table~\ref{tbl:stats} and the stellar mass distribution of cluster galaxies at all redshifts ($z=0.3-1.5$) can be seen in Figure~\ref{fig:massdist}, normalized by the total number of cluster galaxies.

%\emph{Table 1: redshift bins, number of clusters, number of specz and photoz cluster members, breakdown into late/early types, wedge sources?  The field correction gives us a sense of what fraction of these sources are actually cluster members.  Include those numbers in Table 1? }

\begin{table*}
\begin{minipage}[!ht]{\textwidth}
\caption{Cluster statistics.}
\label{tbl:stats}
\begin{tabular}{lccc}
\hline
Redshift & Number of & Number of & Number of  \\
Bin & Clusters & Spectroscopic Redshift Members & Photometric Redshift Members$^a$   \\
\hline
0.3-0.5 & 60 & 160 & 1539  \\
0.5-0.7 & 55 & 112 &  1956 \\
0.7-0.9 & 52 & 24 & 2423 \\
0.9-1.1 & 49 & 20 & 2482 \\
1.1-1.3 & 30 & 58 & 1383 \\
1.3-1.5 & 28 & 47 & 1320 \\
\hline
\end{tabular}
\\
\tiny{$^a$Not corrected for field contamination (see Section~\ref{sec:clusterstacks}).}\\
%\tiny{$^b$AGN classification is determined based on X-ray and IRAC \\
%colors \citep[][]{kir13}. }
\end{minipage}
\end{table*}

Constraining the integral of $P(z)$ provides both an indication of whether a given galaxy is a cluster member or a foreground/background source and a cut on the quality of the photometric redshifts used throughout this study.  We expect that some fraction of the galaxies identified as cluster members will actually be contaminating field galaxies due to the width of the redshift probability distribution functions.  We mitigate this effect on our stacking analysis in two ways: (1) we test the effect of raising the integrated $P(z)$ threshold on our results, and (2) we estimate and subtract the field contamination directly using our field galaxy population.  We find that raising the integrated $P(z)$ threshold has little effect on our overall conclusions and is likely removing real cluster members from our sample.  Our field contamination correction, based on a statistical analysis, is described in Section~\ref{sec:clusterstacks}.  

We also expect some of our galaxies to host AGN.  Using shallow X-ray observations across the field (see Section~\ref{sec:xray}) and IRAC colour selection \citep{ste05, kir13}, we identify AGN in a small fraction of our cluster galaxies, $\sim1-3\%$ from low to high redshift.   More thorough studies of AGN in the ISCS, using deeper X-ray data, have shown that the fraction of AGN in cluster galaxies is as much as 10 per cent at $z>1$, an increase of two orders of magnitude from local AGN fractions \citep{gal09, mar13}.  Given a constraint of $\lesssim10\%$ and the fact that we are primarily probing with the cold dust regime which has been found to be dominated by heating from star formation even in known AGN \citep[e.g.,][]{mul12}, we choose to leave AGN in our sample for our main analysis.  We examine the impact of AGN on our SPIRE stacking analysis separately in Section~\ref{sec:agn}.

\subsubsection{Field Galaxies}
\label{sec:field}

Our field galaxy set is drawn from the Bo\"{o}tes field using the same 4.5$\mu$m-selected galaxy catalogue as the cluster members.   In order to get a clean field sample, we discard any galaxies that are within a radius of 2.5 Mpc and 0.2(1+$z$) in redshift space of known clusters.  We further restrict our field sample by requiring that each galaxy have an integrated $P(z)$ $\geq$ 0.3 at its best-fitting redshift, which places a similar cut in the quality of the photometric redshifts as our cluster member sample. This ensures that we are looking at similar galaxy populations in terms of our ability to assign an accurate photometric redshift.  The stellar mass distribution of our field galaxy sample at all redshifts can be seen in Figure~\ref{fig:massdist}, normalized to the total number of field galaxies.  Though a more careful analysis of the stellar mass function is beyond the scope of this work, we can see that, within the uncertainties, the mass distributions for cluster and field galaxies are similar, with a suggestion of a small difference in the normalized fraction of galaxies in each environment in the low and high mass ends.  We find the same relative mass distributions of cluster and field galaxies when we split out galaxy samples into $z<1$ and $z>1$ bins.

\subsection{Far-Infrared: SPIRE Imaging}
\label{sec:fir}

The Bo\"{o}tes field was observed with $Herschel$ SPIRE \citep[250, 350, and 500$\mu$m;][]{gri10} as part of the $Herschel$ Multi-tiered Extragalactic Survey \citep[HerMES;][]{oli12}.  The observations covered 8 square degrees of the NDWFS/Bo\"{o}tes field, centered on 14:32:06 +34:16:48, which was surveyed with four pointings.  The central two square degrees of the field were then observed with an additional 5 pointings.   We will refer to the smaller, deep area as the ``inner" region and the shallower, wider area as the ``outer" region throughout this work.  In order to optimize these data for point source recovery, we reduced and mosaicked the publicly available AORs using the Herschel Interactive Processing Environment version 7 \citep[HIPE;][]{ott10}, focusing on the removal of striping through high order polynomial baseline removal, the correction of astrometry offsets through the stacking of the positions of known $Spitzer$ Multiband Imaging Photometer (MIPS) 24$\mu$m sources, and the removal of glitches missed by the standard pipeline reduction.  The final maps have a 5$\sigma$ depth in the inner (outer) region of 14mJy (26mJy) at 250$\mu$m.  The confusion noise for SPIRE observations is discussed in \citet{ngy10} and is 5.8$\pm$0.3 mJy at 250$\mu$m.  We generated 5$\sigma$\footnote{confusion noise is not included in the S/N estimates for the catalogue sources} point source catalogs at 250, 350, and 500$\mu$m, both for the original unfiltered maps and after using a matched-filter technique, a method developed to optimize the signal-to-noise ratio (S/N) for confusion-dominated submillimetre maps \citep[see][]{cha11}.  Completeness simulations indicate that our source catalogue is 90$\%$ complete in the inner (outer) region down to 20mJy (33mJy) at 250$\mu$m for the unfiltered map and down to 18mJy (25mJy) for the matched-filter map.   A more detailed description of the reduction of the 250, 350, and 500$\mu$m SPIRE maps, their catalogs, and the completeness tests performed can be seen in Appendix~\ref{appendix:a}.  

The large full width at half-maximum (FWHM) of the SPIRE imaging \citep[18\arcsec.1, 24\arcsec.9, and 36\arcsec.6 at 250, 350, and 500$\mu$m;][]{swi10} presents challenges for both detected sources and for stacking analyses.  Clustering and source blending will result in flux boosting within the large beams, particularly at the longer wavelengths.  For detected sources, we address this by simulating the flux boosting as a function of flux (see Appendix~\ref{appendix:a}).   The bias introduced into SPIRE stacking analyses due to clustering has recently been examined in two studies.  \citet{bet12} found that boosting due to clustering of sources ranges from $\sim7\%$ at 250$\mu$m to $\sim20\%$ at 500$\mu$m for typical galaxy densities in the field, however \citet{vie13} showed that this bias factor increases dramatically with increasing source density and increasing beamsize (see their Figure~4).  In addition, the typical region examined in this study (0.5 Mpc or 1 arcminute radius at $z$=1), will be covered by $\lesssim$2 beams at 500$\mu$m.   For these reasons, we limit our stacking analysis to the 250$\mu$m waveband and apply a correction for clustering bias by determining the baseline signal in the map through a random sampling of pixels both in the field and in the areas of our map which contain clusters (see Section~\ref{sec:clusterstacks}).

\subsection{Mid-Infrared: MIPS Imaging}
\label{sec:mir}

To constrain any evolution in the SEDs of cluster galaxies relative to coeval
field galaxies, we also stack the MIPS AGN and Galaxy Evolution Survey (MAGES;
Jannuzi et al., in preparation)
$70\mu$m images at the positions of our cluster and field galaxies (see Section~\ref{sec:mipsstacking} for details on the stacking of the $70\mu$m images).

MAGES imaged the Bo\"otes field to a depth two times greater than the original
Guaranteed Time Observations (GTO) survey of the Bo\"otes field in each of the three MIPS bandpasses \citep{rie04}.  
The MAGES data also added three additional spacecraft roll angles, which allow 
for improved rejection of $1/f$~noise in the resulting
maps.  The flatter backgrounds in the $70\mu$m~and $160\mu$m~MAGES images
compared to the original survey allow reliable stacking in these bands.

The MAGES data were reduced using the MIPS-GTO pipeline 
\citep{gor05}, and source catalogs
were generated from the resulting image mosaics with D{\lowercase{\sc AOPHOT}} \citep{ste87}.  The MAGES 
point-source catalogs reach $3\sigma$ sensitivities of 0.122, 18.6 and 110
mJy in the 24, 70, and $160\mu$m~images, respectively.

\subsection{X-ray: $Chandra$ photometry}
\label{sec:xray}

%\emph{Describe the available X-ray data used here.}

X-ray data is available across a 9.3 square degree field as part of XBo\"{o}tes, a mosaic of 126 short (5ks) $Chandra$ ACIS-I images covering the entirety of NDWFS \citep{mur05, ken05}.  The XBo\"{o}tes catalogue contains 2,724 point sources with energies of 0.5-7 keV, which is sufficient to detect unobscured moderate to luminous AGN \citep{ran03}.


\section{Stacking Analyses}
\label{sec:stacking}

Stacking is a statistical process by which the signal from multiple individually undetected sources is combined in order to increase the overall S/N and obtain a representative (commonly mean or median) flux density of a population in some waveband \citep[e.g.,][]{dol06, mar09, bet12}. The details of the stacking process depend on the map and the spread in the properties of the population being stacked.  Stacking will allow us to probe much deeper down the infrared luminosity function than requiring detections, as most of the ISCS cluster galaxies will be undetected given the 250$\mu$m flux limit (14mJy or L$_{\rm IR}\sim5\e{11}\Lsun$).  We describe our stacking procedure at 250$\mu$m and our main stacking analysis of cluster and field galaxies in this section.  In addition, we describe stacking procedure at MIPS 70$\mu$m, which will be used to verify our results at 250$\mu$m.

\subsection{Stacking at 250$\mu$m}
\label{sec:spirestacking}

\subsubsection{Procedure}
\label{sec:procedure}

Stacking at 250$\mu$m is performed on the unfiltered map, which has a zero mean and is calibrated in Jy beam$^{-1}$.  The latter fact greatly simplifies the stacking process as the peak pixel value provides the best estimate of the total flux density of a given source at that position (in the absence of clustering or source blending).  The signal of a stack is therefore obtained by combining the pixels in which the sources being stacked are located.  Given that our map has two regions with differing noise properties, we choose to combine the pixel values at the locations of the sources in each stack using a variance-weighted mean.  Stacking tests on fake sources inserted into the map (see Appendix~\ref{appendix:a}), however, show that both variance-weighting and unweighted schemes provide equally good estimates of the true stacked flux.  

The uncertainties associated with each stacked flux density are obtained via the bootstrap method,  during which random subsamples (with replacement) of sources are chosen and re-stacked.  The number of sources in each subsample is equal to the original number of sources in the stack.  This process is repeated 10 000 times in order to determine the representative spread in the properties of the population being stacked.  The bootstrap uncertainty $\sigma_{boot}$ can be expressed by 

\begin{equation}
\label{eqn:boot}
\sigma_{boot} = \frac{\sqrt{\sigma^2_{instr} + \sigma^2_{conf} + \sigma^2_{pop}}}{\sqrt{N_{stack}}}
\end{equation}
where $\sigma_{instr}$ is the instrument noise, $\sigma_{conf}$ is the confusion noise, $\sigma_{pop}$ is the intrinsic spread in the flux density of the population being stacked, and $N_{stack}$ is the number of sources in the stack.  As discussed in \citet{bet12}, though $\sigma_{conf}$ and $\sigma_{pop}$ are most likely not Gaussian, $\sigma_{boot}$ can be approximated as a Gaussian via the central limit theorem given a large number of stacking iterations.  Bootstrapped uncertainties are advantageous as they provide an indication of the scatter in a population, which may include extreme outliers which are otherwise not obvious in a straight measurement of the mean and the standard deviation.  

The process of stacking in general is best understood for sources below the detection limit, where each individual measurement is dominated by Gaussian noise.  We test the contribution from detected sources by matching our cluster members to the 250$\mu$m matched-filter catalogue. The large beamsize and relatively low S/N of the SPIRE observations creates large offsets between the true position of the submillimetre flux and where it is detected in the maps due to random noise peaks.  We characterize these positional uncertainties as part of our completeness simulations (see~Appendix~\ref{appendix:a}) and determine that a search radius of 8'' is appropriate to identify the vast majority of 250$\mu$m counterparts.  We find that $\lesssim$10$\%$ of our cluster members ($r<0.5\,$Mpc) have a 250$\mu$m counterpart within 8'' of their position.  This is not unexpected, given that the deep inner region of the 250$\mu$m map has a flux limit of $14\,$mJy, which corresponds to L$_{\rm IR}\sim5\e{11}\Lsun$ at $z$=1.  A test of random positions across the 250$\mu$m map indicates that we expect a chance encounter with a detected source in an 8'' search radius at a rate of $\sim3\%$.  Given that only a small fraction of our cluster sample is detected at 250$\mu$m, we treat all of our cluster members as undetected and stack them accordingly.  We verify this approach by examining the distribution of flux values that go into each stack, which should be Gaussian due to the noise properties of the undetected sources and have a well-defined mean.

\subsubsection{Stacking Cluster Members}
\label{sec:clusterstacks}

Cluster members identified as described in Section~\ref{sec:membership} are stacked in redshift bins with a width of 0.2 over the redshift range $z=0.3-1.5$ and radial bins as described below.  The mean redshift of each bin is calculated as the mean of the best-fitting redshifts of the constituent galaxies.  As discussed in Section~\ref{sec:redshifts}, to obtain a complete mass-limited sample over our redshift range, we impose a stellar mass limit of $M_{\star} = 1.3\e{10} \Msun$.  

In order for the average flux values of our cluster member stacks to be meaningful, we need to remove any signal that is unrelated to real cluster members.  There are two potential sources of contaminating signal in our stacks: 1) an underlying, baseline signal, mainly due to source blending and clustering, (with a possible minor contribution from dust in the intercluster medium (ICM) contributing a few percent to the IR luminosity \citet{gia08}), and 2) contamination by field galaxies which are mistaken for cluster members due to the width of the photometric redshift probability distribution functions.  

\begin{figure}
\hspace{-5mm}
\includegraphics[scale=0.095, trim=18mm 20mm 20mm 25mm, clip]{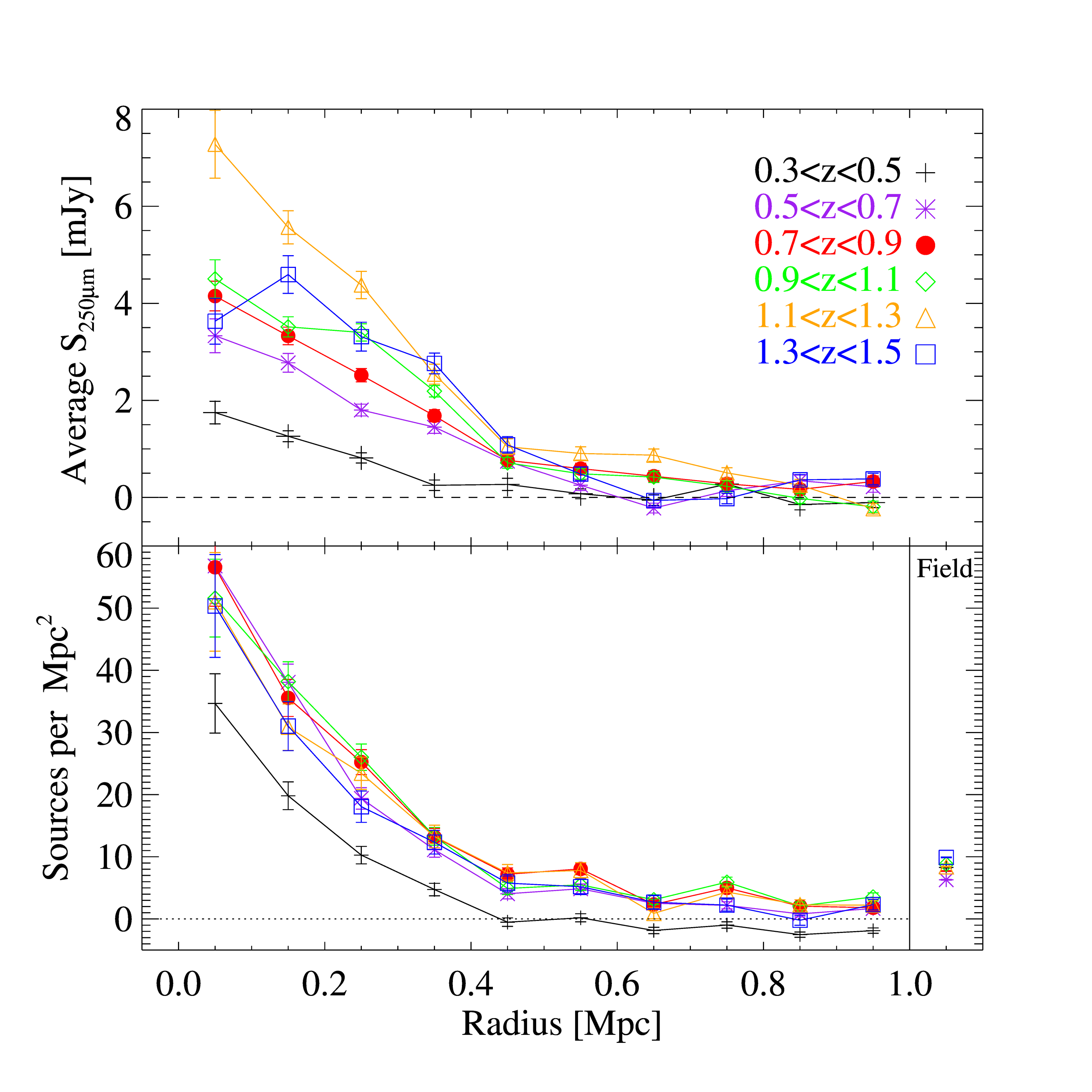}
\caption{(Top) The average 250$\mu$m flux density in randomly-selected pixels as a function of projected cluster-centric radius in redshift bins.  This baseline signal is due to the increased source density toward clusters (resulting in source blending and clustering signal) and must be removed from the stacking signal in the areas of the SPIRE map that have clusters.  (Bottom) The source surface density of cluster members after correcting for field contamination.  The density of cluster members at $r\lesssim$0.5 Mpc dominates over the background field level, while at $r\gtrsim$0.5 Mpc the corrected source density of cluster members is only a small enhancement over the field source density. }
\label{fig:baseline}
\end{figure}

First, we test the 250$\mu$m map for a baseline signal towards the clusters.  SPIRE maps are normalized such that they have a zero mean baseline, which we verify by stacking on 100 000 random pixels across the 250$\mu$m map.  This indicates that there is no overall baseline signal that needs to be removed and boosting from clustering bias of all galaxies across the map is negligible.  The increased source density inherent in the clusters themselves, however, can cause a underlying signal due to source blending and the strong clustering of galaxies in clusters.  To examine this signal, we split the clusters into the redshift bins described above and stack random pixels in projected radial bins originating at the cluster centers.  Figure~\ref{fig:baseline} (top) shows the average 250$\mu$m flux densities recovered from these random stacks as a function of radius and redshift.  At all redshifts, the baseline signal in clusters is strong out to $r=0.5$ Mpc, indicating clustering bias and source blending.  At larger radii, where the number density of cluster members drops (bottom panel), the baseline signal is significantly reduced.  Stacking beyond the virial radius ($\sim1\,$Mpc) recovers no signal.  %This baseline signal will be subtracted from the stacked signal of cluster members in the appropriate radial bins.

In addition to redshift bins, we choose projected radial bins such that we get good number statistics in each cluster galaxy stack.  The baseline signal of cluster galaxies (Figure~\ref{fig:baseline}, {\it top}) as a function of radius suggest a division at $r=0.5\,$Mpc, which is approximately half the virial radius given the expected masses and velocity dispersions of these clusters \citep{sta05, els06, bro06, bro07, eis08, bro11}.    We stack all cluster members in six redshifts bins and two radial bins: $r<0.5$ Mpc and $0.5<r<1\,$Mpc, which we will refer to as the cluster ``core" and ``outskirts'' throughout this work.  We re-calculate the baseline signal as described above for the larger radial bins and subtract the baseline signal from the cluster stacked flux densities in the appropriate redshift/radial bins. 

\FloatBarrier

The second correction is for contamination of the cluster member catalogue by field galaxies.  Due to the nature of our criteria for cluster membership, we expect that some fraction of our cluster members are actually field galaxies which are spatially coincident with one of the ISCS clusters and whose photometric redshift probability distribution function satisfies Equation~\ref{eqn:membership}. Given that the width of a cluster in redshift space will be sharply peaked compared to the cumulative width of the photometric redshift probability distribution functions, this contribution can be determined in a statistical fashion by calculating the ``background" total 250$\mu$m flux per unit area of field galaxies which would satisfy Equation~\ref{eqn:membership} if the cluster was not present.  To accomplish this, we mask out a 2.5 Mpc area around all known clusters within the Bo\"{o}tes field and use the remaining area to identify galaxies which have an integrated $P(z)\geq0.3$ at discrete redshifts, ranging from $z=0.3-1.5$ in steps of 0.05.  The galaxies which satisfy integrated $P(z)\geq0.3$ at each redshift step are stacked to determine the mean flux level of field galaxies, $\avg{S_{fc}(z)}$, which we  additionally smooth with a boxcar filter with a width of 0.1 to remove noise introduced by the binning in redshift space.  Multiplying by the number of field galaxies per unit area, $\Sigma_{fc}(z)$, we find the total 250$\mu$m flux per unit area that we can expect to contaminate our cluster stacks.

\begin{figure}
\centering
\hspace{-10mm}
\includegraphics[scale=0.095, trim=0 10mm 30mm 0, clip]{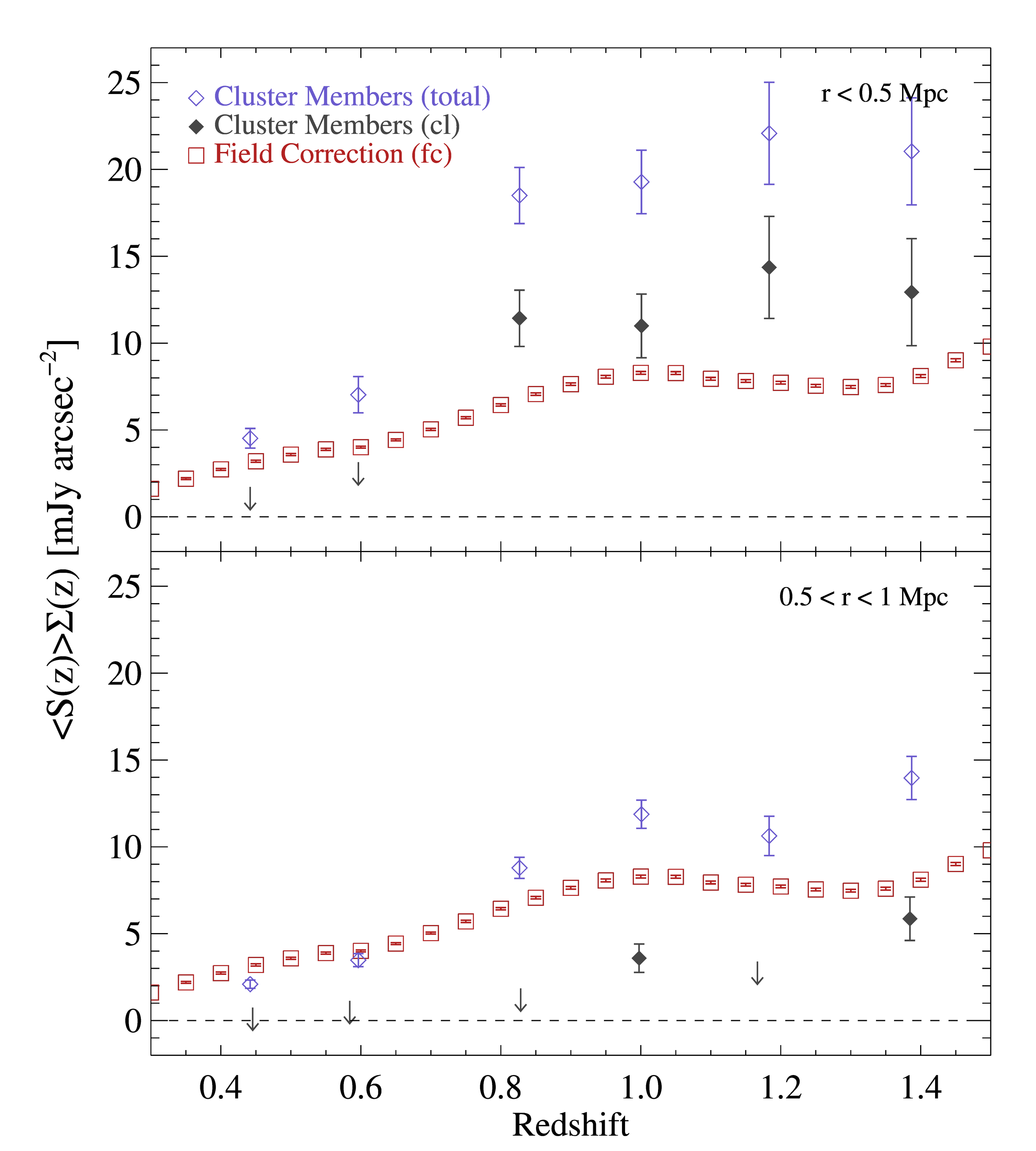}
\caption{The total 250$\mu$m flux per unit area as a function of redshift for cluster members  and field galaxies in radial bins $r<0.5\,$Mpc (top) and $0.5<r<1\,$Mpc (bottom).  This quantity is obtained by multiplying the average stacked flux density, $\avg{S(z)}$ with the number density of sources, $\Sigma(z)$.  The blue, open diamonds are the field-contaminated cluster galaxy stacks (after correction for baseline signal due to source blending/clustering), denoted ``total" in Equations~\ref{eqn:fc}-\ref{eqn:numdensity}.  The black, filled diamonds show the total 250$\mu$m flux per unit area of cluster galaxies denoted ``cl'', after both baseline and field contamination corrections have been applied.  The red, open squares indicate the total flux per unit area of field galaxies which satisfy Equation~\ref{eqn:membership}, denoted ``fc".  No stacked signal above the field contamination was detected for cluster galaxies at $r>1$~Mpc in any of the redshift bins.  Upper limits are 3$\sigma$.}
\label{fig:fluxperarea}
\end{figure}

\begin{figure}
\centering
\hspace{-10mm}
\includegraphics[scale=0.11, trim=23mm 0 0 0, clip]{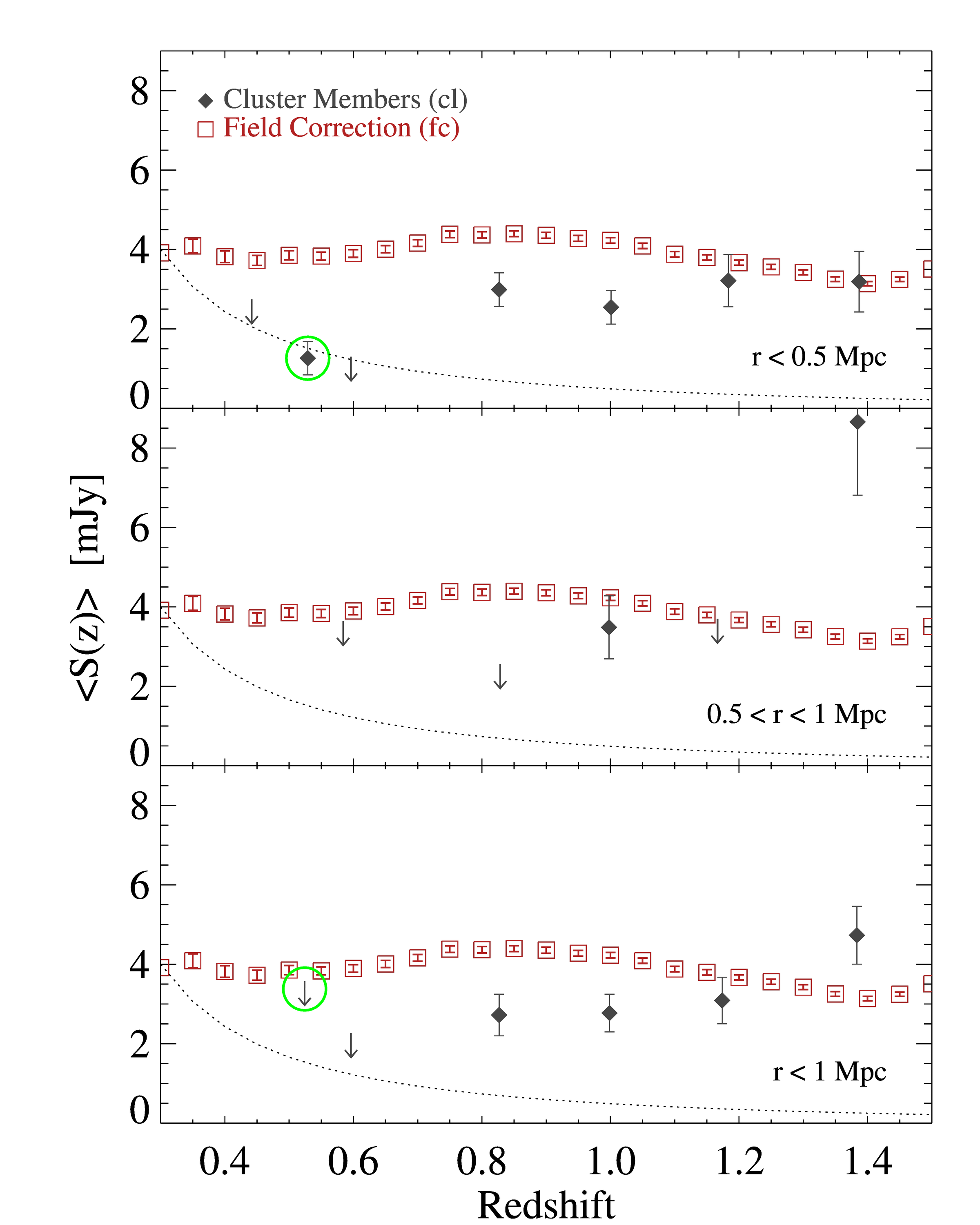}
\caption{The average stacked 250$\mu$m flux density for cluster members after baseline and field contamination corrections (cl; black diamonds) as compared to field galaxies (fc; red squares) which satisfy Equation~\ref{eqn:membership} as a function of redshift.  The top panel shows cluster members for $r<0.5\,$Mpc, the middle panel for $0.5<r<1\,$Mpc and the bottom panel combines the two for an $r<1\,$Mpc bin to maximize detections.  In addition, we combine the two lowest redshift bins into a bin spanning $z=0.3-0.7$ (indicated by green circles).  The average 250$\mu$m flux densities of field galaxies which satisfy Equation~\ref{eqn:membership} are consistent with the average fluxes of all field galaxies in our mass-limited sample (Section~\ref{sec:field}).  The dotted lines show the 250$\mu$m $k$-correction for a typical dusty, star-forming galaxy of constant luminosity  and normalized to the field level at $z=0.3$.  Upper limits are 3$\sigma$.  Some lower redshift bins are poorly constrained and their upper limits are outside the plot ranges.}
\label{fig:stackedflux}
\end{figure}

 The field correction is applied by subtracting the total flux per unit area of field galaxies (Figure~\ref{fig:fluxperarea}, red squares) from the total flux per unit area of our contaminated cluster stacks (Figure~\ref{fig:fluxperarea}, blue diamonds) via:  

\begin{equation}
\label{eqn:fc}
\avg{S_{cl}(z)}\Sigma_{cl}(z) = \avg{S_{total}(z)}\Sigma_{total}(z) - \avg{S_{fc}(z)}\Sigma_{fc}(z)
\end{equation}
where $z$ is the mean redshift of cluster members in a given bin, $\avg{S_{total}(z)}$ and  $\Sigma_{total}(z)$ are the stacked fluxes and the number of sources per unit area in the contaminated cluster stacks, $\avg{S_{cl}(z)}$ is the true flux density of cluster galaxies and $\Sigma_{cl}(z)$ is found via

\begin{equation}
\label{eqn:numdensity}
\Sigma_{cl}(z) = \Sigma_{total}(z) - \Sigma_{fc}(z) .
\end{equation}

The field corrected total 250$\mu$m flux per unit area of cluster galaxies can be seen in Figure~\ref{fig:fluxperarea} as the filled black points for the cluster cores ($r<0.5\,$Mpc; {\it top}) and outskirts ($0.5<r<1\,$Mpc; {\it bottom}).  In the cluster cores, the corrected total 250$\mu$m flux per unit area exceeds that in the field at high redshift ($z>0.8$).  In the outskirts, only two of the redshift bins are detected at $\geq3\sigma$; however, the highest redshift bin shows that the total flux per unit area of cluster galaxies is approaching the level of the field.  This is significant given that the number of cluster galaxies per unit area (see Figure~\ref{fig:baseline}), corrected via Equation~\ref{eqn:numdensity}, is only a small enhancement over the field source density at these redshifts, indicating that the average activity in cluster members will be higher than in the field in this high redshift bin (see Figure~\ref{fig:stackedflux}).

The corrected mean flux density of cluster galaxies, after the baseline and field corrections described above are applied,  can be seen in Figure~\ref{fig:stackedflux} (black diamonds) compared to the average flux of field galaxies per redshift (red squares) for $r<0.5\,$Mpc (top) and $0.5<r<\,1$Mpc (middle).  No stacked signal above the field was found at $r>1\,$Mpc. Four out of six redshift bins are detected at $\geq3\sigma$ for $r<0.5\,$Mpc and two out of six are detected for $0.5<r<1\,$Mpc.   In order to maximize the number of detected stacked signals we have to work with, we make two changes for the subsequent analysis.  1) we combine the two lowest redshift bins, and 2) we combine the core and outskirts bins into one ``core+outskirts" $r<1\,$Mpc radial bin (Figure~\ref{fig:stackedflux}, bottom), which we will compare with the $r<0.5\,$Mpc radial bin.  The two radial bins are combined as a weighted mean after applying the baseline correction.  To verify that these trends are not a product of our binning scheme, we shifted the redshift bins by 0.1 and re-stacked and re-corrected our cluster galaxy stacks.  We find that these trends are robust against the exact redshift bins chosen. 

The shape of the average 250$\mu$m flux of field galaxies in Figure~\ref{fig:stackedflux} is relatively flat, which reflects that the average infrared luminosity of field galaxies is increasing with redshift, compensating for the $k$-correction (dotted curve).  It should be noted that there are several submillimetre emission lines which will be sampled by the 250$\mu$m band over this redshift range.  The brightest, CII (rest-frame 158$\mu$m), has been measured to contribute $\sim4$ per cent to the 250$\mu$m flux for $z<1$ for a typical submillimetre galaxy (SMG) and may contribute more in sub-ULIRG galaxies, though this has not been well quantified to date \citep{sma11}.  These emission lines may contribute to the wiggles in the 250$\mu$m flux as a function of redshift for the field.  The (corrected) average 250$\mu$m flux of cluster galaxies, on the other hand, clearly rises as a function of redshift.  We examine these results in terms of physical properties in Section~\ref{sec:results}.

\subsubsection{Stacking Field Galaxies}
\label{sec:fieldstacks}

We stack field galaxies in the Bo\"{o}tes field, the selection of which is described in Section~\ref{sec:field}, in two ways.  First, we stack them in redshift bins with width 0.1 to take full advantage of the large numbers of field galaxies at our disposal in order to examine the evolution of their infrared properties as a function of redshift.  Second, we bin them in the same redshift bins as our cluster galaxies for a direct comparison.  For the latter,  we take $N_{stack}$ (see Equation~\ref{eqn:boot}) from the corresponding cluster stack rather than the total number of field galaxies in the stack, which is typically an order of magnitude larger than the number of cluster members.  This will provide comparable uncertainties.  We find that the average 250$\mu$m flux densities of field galaxies stacked in this way are consistent with the values found for the field correction.  We test that this holds even if we remove the restriction on the integrated $P(z)$ for the field stacks.  This indicates that the cluster membership criteria does not introduce a bias based on the restriction of the integrated $P(z)$ parameter at a given redshift.

\subsubsection{Projection Effects and Verifying the Baseline Correction}
\label{sec:verifying}

In this section, we briefly discuss (i) projection effects due to selecting cluster members based on their 2D cluster-centric radius 
%(ii) the level of field contamination in our cluster stacks which has been removed, 
and (ii) our tests to verify our baseline correction procedure. 

\begin{enumerate}
\item Since we are using projected cluster-centric radii, we are stacking cluster galaxies in cylinders rather than spheres and will suffer some contamination to our signal from projection effects.   A recent study by \citet{nob13} examined contamination due to projection effects by separating infalling galaxies from older cluster populations using caustic diagrams.  They found that recently accreted, star forming galaxies contaminate at all projected radii and that this effect may be responsible for recent studies claiming no environmental dependence for star forming galaxy properties as a function of radius. As our radial bins are quite large, we expect our susceptibility to this to be minimized, however, we can quantify these effects in the following way.  For the $r<1\,$Mpc bin, we argue that projection effects are not significant as we found no stacked signal above our field contamination outside 1~Mpc.  The $r<0.5\,$Mpc bin most likely contains some signal from cluster galaxies at larger radii.  Stacking at $0.5<r<1\,$ Mpc found that only 2/5 redshift bins are detected, with the strongest signal in the outskirts at $\avg{z}$=1.4.   By comparing the field-corrected source density of cluster members in the core and outskirts in our highest redshift bin and using the average stacked flux from each to determine their relative contribution to the total flux and source densities, we estimate that the outskirts are contributing $\sim30\%$ of the average flux in the $r<0.5\,$Mpc $\avg{z}$=1.4 bin.   

%\item Using Equation~\ref{eqn:numdensity}, we quantify the percentage of contaminating field galaxies in our total cluster stacks as a function of redshift.  The fraction of contaminating field galaxies is $\sim30\%$ ($\sim50\%$) for all bins except the lowest and highest redshift bins for $r<0.5\,$Mpc ($r<1\,$Mpc).  For our lowest redshift bin ($\avg{z}$=0.5), the fraction of contaminating field galaxies is $\sim40\%$ ($\sim70\%$) and, for the highest ($\avg{z}$=1.4), $\sim40\%$ ($\sim60\%$) for $r<0.5\,$Mpc ($r<1\,$Mpc).

\item We test our general stacking technique and method for extracting the baseline signal due to galaxy clustering and source blending by re-stacking our $r<0.5\,$Mpc cluster members and field galaxies using S{\lowercase{\sc IMSTACK}} from \citet{vie13}, which was developed to account for clustering bias in stacking analyses.  We find that S{\lowercase{\sc IMSTACK}} yields consistent results with our stacking method, indicating that any clustering bias in the full field population is negligible, as our baseline test in the field determined, and that we are correctly removing the signal from clustering bias in our $r<0.5$ cluster stacks.  The \citet{vie13} code is designed to stack  populations of galaxies with similar clustering properties and so we do not test our outer radial bin, as they mix cluster and field galaxies in similar proportions.  
\end{enumerate}

%The \citet{vie13} code is currently untested at the source densities of our clusters and the mixing of cluster and field populations may have unforseen effects on their stacking method.  As such, we do not run our cluster stacks using their code.  \emph{(Note to coauthors:  I did actually run their code for the cluster members and found good agreement with my results for the $r<0.5$ Mpc bin, but not for the $0.5<r<1$ Mpc bin.  We are pretty confident in our baseline correction scheme and wonder if their code is unable to handle the mixture of field and cluster populations at large cluster radii.  This remains an unresolved issue for now.)}.s

%Using the total number of sources per area toward the clusters in each redshift bin and the number of field galaxies per area which satisfy Equation~\ref{eqn:membership} at the average redshift of a given bin, we determine that the fraction of contaminating field galaxies is $\sim$30$\%$ ($\sim$52$\%$) for all bins except the lowest and highest redshift bins for $r<0.5$ Mpc ($r<1$ Mpc).  For our lowest redshift bin ($\avg{z}$=0.4), the fraction of contaminating field galaxies is $\sim$51$\%$ ($\sim$88$\%$) and, for the highest ($\avg{z}$=1.4), $\sim$42$\%$ ($\sim$63$\%$) for $r<0.5$ Mpc ($r<1$ Mpc).

\subsection{Testing the Contribution of Active Galactic Nuclei at 250$\mu$m}
\label{sec:agn}

As discussed in Section~\ref{sec:membership}, a small fraction ($\leq10\%$) of the galaxies in our field-contaminated cluster stacks are expected to host AGN, which we expect to have a minimal contribution to the cold dust regime as probed at 250$\mu$m over our redshift range. To confirm this is true for the bright AGN we can detect across our galaxy samples, we remove all galaxies which 1) have an X-ray detection and/or 2) fall in the IRAC colour selection ``wedge'' for AGN as described in \citet{kir13} and repeat our stacking analysis as described in Sections~\ref{sec:clusterstacks} and \ref{sec:fieldstacks}.  We find that removing these AGN makes no statistically significant difference in the measured stacked fluxes for either our cluster or field galaxy samples.

%\emph{\color{red}I have redone the stacking analysis with all X-ray and IRAC selected AGN taken out and found no difference from the general results (Figure 4), which confirms that we are not suffering from AGN contamination.  This isn't surprising, but it is nice to confirm.  I am in the process of stacking the AGN themselves, but it is tricky due to low number statistics  and I may not find anything useful.  }

\subsection{Stacking at 70$\mu$m}
\label{sec:mipsstacking}

The MAGES 70$\mu$m flux maps differ from the 250$\mu$m maps described above in several respects, including larger spatial variations in sensitivity, the units of the image mosaic, and the relative importance of confusion noise. As a result, we treat the 70$\mu$m stacks slightly differently than the 250$\mu$m stacks. In this section, we describe the procedures we used to stack the MAGES 70$\mu$m image for both the field and cluster galaxy samples and the corrections that we applied to photometry measured from stacks of cluster galaxies.

\subsubsection{Procedure}\label{sec:Proc70}
The MIPS $70\mu$m bandpass is more sensitive to the presence of warm and hot
dust than the SPIRE $250\mu$m bandpass.  This is especially true in our
higher-$z$ bins, in which the 70$\mu$m band probes rest-frame wavelengths
$\lambda\lesssim 30\mu$m.  As a result, the L$_{\rm IR}$ inferred at 
$\lambda_{obs}=70\mu$m can be more strongly influenced by a small population
of galaxies with unusually warm dust than can L$_{\rm IR}$ inferred at 250$\mu$m.

Since detected sources contribute more at 70$\mu$m than at 250$\mu$m where confused sources dominate,  we use a residual image for stacking to avoid contribution from the wings of unrelated bright sources near the target positions.  The 70$\mu$m residual image is constructed by point spread function (PSF) subtracting all sources detected at 5$\sigma$ significance from the 70$\mu$m science image using D{\lowercase{\sc AOPHOT}}.  Stacked images constructed from the residual image yield a flatter background that is consistent with the intrinsic background in the 70$\mu$m science image.  This allows more reliable photometry of the stacked images; however, our use of the residual image requires that we add back the flux from target positions with detected 70$\mu$m counterparts to the flux measured from the stacked image.  We determine the mean flux of galaxies in each redshift bin as,
%\begin{equation}\label{eqWeightFlux}
%f_{\nu}(70\mu m, z) = \frac{(N_{obj}-N_{det})f_{\nu,stack} + N_{det}\langle f_{\nu,det}\rangle}{N_{obj}}
%\end{equation}
\begin{equation}\label{eqWeightFlux}
S^{70\mu\mathrm{m}} = \frac{(N_{stack}-N_{det})S_{stack} + N_{det}\langle S_{det}\rangle}{N_{stack}}
\end{equation}
where $S_{stack}$ is the flux measured from the stacked image, and
$\langle S_{det}\rangle$ is the mean flux of detected galaxies.  The numbers of sources $N_{stack}$ and $N_{det}$ indicate the total number of galaxies in the appropriate redshift bin and the number of 
targets in the stack with detected counterparts, respectively.  We tested whether spatial variations in the uncertainties of individual pixels require variance weighting in the mean-combined stacks and found that weighting the stacked images makes no difference in our ability to recover the mean fluxes of galaxies in our source lists.  In order to determine the uncertainties associated with each stack, we also generate 2500 mean-combined, bootstrap-sampled image stacks from the residual image.

%We built 2500 mean-combined, bootstrap-sampled image stacks from the MAGES 70$\mu$m residual image.  To construct the $70\mu$m residual image, we PSF-subtracted all sources detected at $5\sigma$ significance from the $70\mu$m science image using DAOPhot.   Since detected sources contribute more at 70$\mu$m than at 250$\mu$m where confused sources dominate, our use of the residual image allows us to avoid contributions from the wings of unrelated bright sources near the target positions.  

%In contrast with the mean-combined stacks, we built the median stacks directly from the 70$\mu$m science image.  The nature of the median combination means that the background in the stacked images remains insensitive to the small number of  target positions for which a given sky pixel includes a detected source.  As a result, we do not need an analogue of Equation~\ref{eqWeightFlux} to add the  fluxes of detected sources to the fluxes measured from the median stacks.

%We do not re-sample the image pixels in order to place the nominal position of each source exactly at the center of the stacked images.  The FWHM of the MIPS 70$\mu$m PSF ($16^{\prime\prime}$) is large compared to the maximum error that can be introduced by not re-sampling the pixels ($0.5~{\rm pixels} = 2\farcs{47}$), and we avoid the additional noise that would be introduced in the re-sampling process.

We use aperture photometry to measure fluxes from the mean-combined images.  We measure fluxes in radii of $16\arcsec$, equal to the FWHM of the 70$\mu$m PSF, and we use annuli extending from $r=18\arcsec$ to $r=39\arcsec$ to measure the sky flux.  The measured fluxes are aperture-corrected to $r= \infty$ using a multiplicative factor of 1.21\footnote{MIPS Instrument Handbook \url{http://irsa.ipac.caltech.edu/data/SPITZER/docs/mips/mipsinstrumenthandbook/}}.  The uncertainties are obtained from the RMS dispersion about the mean bootstrapped flux.  
%The fluxes we use inour analysis are listed in Table \ref{tabMipsFluxes}.  We average the fluxes given by Eq. \ref{eqWeightFlux} to construct the fluxes listed in Table \ref{tabMipsFluxes}, and the quoted uncertainties are given by the RMS dispersion about the mean bootstrapped flux.

%The MAGES 70$\mu$m images include significant spatial variations in the flux uncertainties of individual pixels.  These variations are due to a combination of Poisson noise in very bright pixels and variations in the number of overlapping pixels from the original data available in the construction of the final map.  We explored whether these variations require weighting of individual pixels by their uncertainties ($1/\sigma^{2}$).  To test the need for weighting in our analysis, we used DAOPhot to add false sources to our 70$\mu$m mosaics and stacked at the positions of these false sources both with and without inverse-variance weighting.  We found no significant difference in the fluxes returned by the two methods, and the impact of weighting was similarly minimal in both mean- and median-combined stacked images.  We concluded that weighting the images would make no difference in our ability to recover the mean fluxes of galaxies in our source lists.

\subsubsection{Stacking Cluster Members at 70$\mu$m}\label{secClustStack70}
The most important difference between the analysis applied to stacked images
at 70$\mu$m and 250$\mu$m is the absence of an additional baseline correction to the
70$\mu$m fluxes.  The requirement to use aperture photometry to measure
70$\mu$m fluxes, as opposed to the direct measurement of flux from the 
brightest pixel in the 250$\mu$m images, means that the fluxes 
%listed in Table \ref{tabMipsFluxes} 
have already been corrected for the elevated background
in the clusters.  No additional background correction is required.  The field correction is applied to the 70$\mu$m fluxes as described in Section~\ref{sec:clusterstacks}.

%As a check, we constructed 70$\mu$m stacked images at the same random positions used for the 250$\mu$m stacks, as described in Section~\ref{sec:clusterstacks}.  The mean fluxes in the images stacked at random positions within the clusters are elevated relative to the background of the 70$\mu$m science image, and the trend with redshift is qualitatively similar to that shown in Figure 1.


\section{Stacking Results}
\label{sec:results}

%\emph{(Note to co-authors: The figures in this section, which are somewhat redundant looking, are not finalized and may not all appear in the final paper.  The formatting and section titles are likewise not finalized.  Suggestions are welcome. )}

\subsection{Deriving the Total L$_{\rm IR}$, SFRs, and SSFRs from Stacking at 250$\mu$m}
\label{sec:sfr}

Using the stacked 250$\mu$m flux densities, we infer the average physical properties of our cluster galaxies and field galaxies as a function of redshift and cluster radius, including the total infrared luminosity (L$_{\rm IR}$), defined over the rest-wavelengths 8-1000$\mu$m, star formation rate (SFR), and specific star formation rate (SSFR=SFR/M$_{\star}$).  Over our redshift range, the 250$\mu$m waveband probes the far-infrared portion of a galaxy SED, which is dominated by emission from cold dust  heated by star formation. We derive these quantities by comparing to an empirical template developed in \citet{kir12}.  This template was formulated using a sample of star forming galaxies at $0.4<z<1.4$ (L$_{\rm IR}\sim10^{11}\Lsun$) selected at 24$\mu$m and identified as star forming through IRS spectroscopy.  Using deep $Herschel$ imaging over the 100-500$\mu$m wavelength range, the dust properties of the template were modelled using a two-component blackbody.  

The choice to represent the average properties of star forming galaxies with one template is appropriate given that we are measuring the average flux of similar populations and is consistent with our goal of comparing the average star formation properties of cluster galaxies versus field galaxies as a function of redshift.  Template-to-template variations in the far-infrared will be driven by differences in the dust properties of star forming galaxies, which will, to first order, contain a cold dust component from star formation heating of the interstellar medium (ISM) and warm dust components originating from young star forming regions or AGN emission.  In terms of the SED, these details determine the location of the peak of the dust emission and the shape of the Rayleigh-Jeans tail.  Before $Herschel$, only templates from local starbursting galaxies were available for fitting high redshift star forming galaxies; however, these local templates often lacked data spanning 160-850$\mu$m and so had difficulty constraining  cold dust emission.  Multiple studies have shown that high redshift star forming galaxies at the LIRG and ULIRG level may have colder dust than their local counterparts, making the application of local templates to high redshift galaxies problematic \citep{row04, row05, pop06, sym09, sey10, muz10, elb10, nor10, ruj11}.  Using an empirical template with well-sampled far-infrared data and based on high redshift galaxies mitigates some of these concerns; however, we must still address whether it is appropriate to apply one template over the redshift range in this study.   \citet{che13} examined the dependence of the scatter in $S_{250}$/L$_{\rm IR}$ on differing SED shapes for a sub-set of star-forming $z\sim1$ galaxies from \citet{kir12}, finding that the deviations in the far-infrared SED shape are reasonably small and the estimation of L$_{\rm IR}$ from the monochromatic 250$\mu$m flux is appropriate for representative star-forming populations.  In addition, \citet{hwa10} examined the dust properties of galaxies out to $z$=3 using $Herschel$ Photodetector Array Camera and Spectrometer (PACS) and SPIRE data and found the relation between the total infrared luminosity and dust temperature to be fairly constant at sub-LIRG luminosities, with a small rise of $\sim5$K in galaxies with $10^{11} < \rm{L} < 10^{12}\, \Lsun$.  Based on previous studies \citep{bro13} and the rate of detection of cluster members in our shallow SPIRE data, we expect the typical luminosities of our galaxies to be significantly $<10^{12}\Lsun$.  The \citet{hwa10} results then suggest that our galaxies should have fairly consistent dust properties over the redshift range probed.   While a different choice in templates may affect the absolute level of the physical properties inferred in this study, it should not affect the differences we quantify between cluster and field galaxies, if the templates are applied consistently.  We further test this assumption by stacking the same galaxies at 70$\mu$m in Section~\ref{sec:environment}.  A comparison between the \citet{kir12} LIRG template and the commonly used \citet{cha01} templates can be found in \citet{kir12}.

In the following analysis, we estimate the total L$_{\rm IR}$  by normalizing the \citet{kir12} SED template to the stacked 250$\mu$m flux densities of our cluster and field galaxy samples.  The error associated with the SED template is 40$\%$, which accounts for the spread in the SEDs of high redshift star forming galaxies.   From the L$_{\rm IR}$, we obtain the SFR via the \citet{mur11} relation

\begin{equation}
\mathrm{SFR [\Msun yr^{-1}}] = 1.47\e{-10} \mathrm{ L_{IR} }[\Lsun]
\end{equation}
which assumes a Kroupa IMF \citep{kro01}, providing a similar normalization to the Chabrier IMF used to calculate the stellar masses.  Specific star formation rates are calculated from the average SFR (obtained from stacking) multiplied by the number of sources stacked divided by the sum of the masses of the galaxies in the stack.  The number of sources and total mass are corrected for field contamination in the same manner as the stacked fluxes, by subtracting the total number or mass per unit area of field galaxies from the field-contaminated cluster samples.  The error on the total mass in any given bin is determined by bootstrapping.

\subsection{Evolution of Star Formation in Clusters and Field Galaxies with Cosmic Time}
\label{sec:environment}

In this section, we examine the average dust-obscured star formation activity in cluster galaxies as a function of environment by comparing cluster galaxies in two projected radial bins, $r<0.5\,$Mpc (core) and $r<1\,$Mpc (core+outskirts), to our field galaxy sample.  We examine trends in physical galaxy properties as a function of cosmic time and redshift, to better connect our results to the characteristic time-scales of different cluster processes.

In the cluster cores, the average L$_{\rm IR}$ and SFR of our mass-limited sample of cluster galaxies (Figure~\ref{fig:masslimited05}, left, black diamonds) rises rapidly as a function of redshift, drawing even with the field activity (blue circles) at $z\gtrsim1.2$.  Our low redshift bin ($\avg{z}=0.5$) has an average $\textup{SFR}\sim$few $\Msun$ yr$^{-1}$, quenched to $\sim30\%$ of the field level.  In our highest redshift bin, $\avg{z}=$1.4, the average cluster galaxy SFR is consistent with the field at $\sim30\Msun$ yr$^{-1}$.  The $\avg{\textup{SSFR}}$ (Figure~\ref{fig:masslimited05}, right) shows a similarly rapid trend, with the ratio of SSFR in the clusters to the field doubling over this redshift range.   The clusters are suppressed to $\sim30\%$ of the field level at $\avg{z}$=0.5 versus $\sim75\%$ of field level at $\avg{z}=1.4$.   Over our redshift range ($z$=0.3-1.5 or $\sim6$ Gyr), the SSFR in cluster galaxies increases an order of magnitude from  $\sim0.05$ to 0.5 Gyr$^{-1}$.  

Including all cluster galaxies out to $r<1\,$Mpc (Figure~\ref{fig:masslimited1}) dramatically raises the average SFR in the highest redshift bin to $\sim60\,\Msun$ yr$^{-1}$.   In the radial bin $0.5<r<1\,$Mpc (not shown), the $\avg{z}$=1.4 bin is detected at the 5$\sigma$ level with $\avg{\textup{SFR}}\sim90\,\Msun$ yr$^{-1}$,  three times the $\avg{\textup{SFR}}$ in the cluster cores and the field level, with a $\avg{\textup{SSFR}}\sim2$ Gyr$^{-1}$. 

As a check of our measured average field SFR, we compare our field values to 250$\mu$m stacks of K-band selected field galaxies from the UKIRT Infrared Deep Sky Survey \citep[UKIDSS; ][]{law07} in the Ultra-Deep Survey (UDS).  This field galaxy sample extends down to the same stellar mass limit as used in this work and was stacked using S{\lowercase{\sc IMSTACK}} \citep[M. Viero, private communication;][]{vie13}.  We convert the average 250$\mu$m of the UDS sample into a L$_{\rm IR}$ and SFR as described in Section~\ref{sec:sfr} and the results are in good agreement with our field values (Figures~\ref{fig:masslimited05}-\ref{fig:masslimited1}, yellow squares).

\subsubsection{Evolution as a Function of Cosmic Time}
\label{sec:time}

In order to quantify the evolution of the average SFRs and SSFRs for galaxies in clusters versus the field, we fit both the cluster galaxy stacks  (Figure~\ref{fig:masslimited05}, black diamonds) and high resolution field galaxy stacks (small blue circles) with a function of the form $y=\beta \textup{e}^{\alpha t}$, where $t$ is cosmic time.  The fits were performed using M{\lowercase{\sc PFIT}} \citep{mark09}.  Table~\ref{tbl:fit} provides a summary of the fit coefficients, where the coefficient uncertainties are the 1$\sigma$ errors from the covariance matrix as determined by M{\lowercase{\sc PFIT}} and the reduced $\chi^2$ values indicate the goodness-of-fit. The average SFRs of cluster galaxies is decreasing with time as  $\avg{\textup{SFR}}_{\textup{cl}}$ $\propto$ $\textup{e}^{(-0.66\pm0.08)t}$ for the cluster cores versus  $\avg{\textup{SFR}}_{\textup{field}}$ $\propto$ $\textup{e}^{(-0.42\pm0.005)t}$ in the field.  These correspond roughly to e-folding times of 1.5 and 2.4 Gyr, with the star formation in cluster galaxies decreasing $\sim2$ times faster than the field. This e-folding time for field galaxies is consistent with that found to be the median H$_2$ consumption time for local spiral galaxies \citep{big11}.  Following this evolution, the average cluster galaxy has SF on par with the average field galaxy at $z\gtrsim$1.2. The $\avg{\textup{SSFR}}$ does not quite draw even with the field at the highest redshift that we probe in this study, which may indicate a difference in the stellar mass distributions between cluster and field galaxies, as is hinted at in our stellar mass distributions for cluster and field galaxies (Figure~\ref{fig:massdist}) and was measured in clusters at $z\sim1$ in \citet{van13}.  We note, however, that the evolution in the $\avg{\textup{SSFR}}$ with cosmic time is consistent within the errors with that of the $\avg{\textup{SFR}}$ and statistically distinct from the evolution of star formation in field galaxies.  This indicates that differences in the stellar mass distributions between cluster and field galaxies cannot be wholly responsible for driving  these trends.  The fit to $y=\beta \textup{e}^{\alpha t}$ for cluster galaxies at $r<1\,$Mpc (core+outskirts) is less well constrained, due to the lack of a $\geq3\sigma$ detection in the lowest redshift bin, but still shows a  significantly faster than the decline in the field galaxy population with $\avg{\textup{SFR}}_{\textup{cl}}$ $\propto$ $\textup{e}^{(-0.76\pm0.10)t}$.
%The evolution for the $\avg{\mbox{SSFR}}$ is comparable to that of the $\avg{\mbox{SFR}}$ (see Table~\ref{tbl:fit} for summary of all fit coefficients), indicating that any differences in the mass distributions of cluster and field galaxies cannot be driving these trends. \emph{(is this an accurate or even meaningful thing to say?  The rise in SSFR is statistically different from that of the SFR…could this be the result of the changing fractions of SF vs passive galaxies ala the Muzzin paper?)}}

 The reduced $\chi^2$ values for the fit to the high resolution field galaxies stacks indicate that a single exponential function is not a good fit to the data.  Fitting two exponential functions to the field galaxies with an break at $z\sim0.8$ greatly improves the goodness-of-fit and we find that the best-fitting $\avg{\textup{SFR}}$ slopes are significantly different, with $\alpha=-0.53\pm$0.01 at $z<0.8$ and $\alpha=-0.28\pm$0.01 at $z>0.8$. This break is reminiscent of the differential ramp up of LIRGs and ULIRGs in the field with time and the general form of the star formation rate density of the Universe \citep[see][]{mur11a, mag13}.  Though we are unable to repeat this analysis for our cluster sample due to poor resolution in the cluster stacks, we note that the cluster galaxy evolution is still distinct from the field at both low and high redshift. A more in-depth look at the evolution of field galaxies as a function of cosmic time is reserved for a future paper.

%This choice of function is somewhat arbitrary and, though it provides a useful quantification of the general trend for comparison, does not necessarily describe the evolution of any given cluster, for which stochastic effects such as group accretion may play a major role.   A summary of the fit coefficients can be seen in Table~\ref{tbl:fit}.  We find that the cluster galaxies have $\avg{SFR}_{cl}\propto$(1+$z$)$^{5.7\pm0.9}$ versus $\avg{SFR}_{field}\propto$(1+$z$)$^{3.9\pm0.6}$ in the cluster cores ($r<$0.5Mpc).  The evolution of $\avg{SSFR}$ is comparable to that $\avg{SFR}$ for both the clusters and the field (Table~\ref{tbl:fit}), indicating that it is not the different mass distributions between cluster and field galaxies which is driving this trend.  When the cluster outskirts are included ($r<$1Mpc), we fit slightly steeper slopes, consistent with the increased activity seen in the $\avg{z}$=1.4 bin (Figure~\ref{fig:masslimited1}).  This effect is not well constrained in these fits, given our data and the simple functional form we are assuming.

\subsubsection{Evolution as a Function of Redshift}
\label{sec:redshift}

Multiple studies have examined cluster properties, such as the star-forming galaxy fraction, number of LIRGs, and total SFR per halo mass \citep[e.g.,][]{bai09, hai09, pop12, web13}, and quantified their evolution as a function of redshift.  Though the cluster properties, quantities measured, and sample selection vary greatly between cluster studies, including this work, it is instructive to assume that all of these quantities are related on some level.  As such, we compare our average SFR as a function of redshift by fitting the commonly adopted form $y=y_0(1+z)^{n}$ to the cluster members and high resolution field stacks, as above.  

In the cluster cores ($r<0.5\,$Mpc), we find that the evolution of the average SFR goes as $n=5.6\pm$0.6, while in the core+outskirts ($r<1\,$Mpc), $n=5.9\pm$1.0.  This is statistically distinct from the field, where $n=3.9\pm0.04$.  The evolution of the SSFR is similar, with $n=5.1\pm$0.7 (core) and $n=5.3\pm$1.1 (core+outskirts), compared to $n=4.0\pm0.05$ for field galaxies.  The coefficients and their reduced $\chi^2$ values are summarized in Table~\ref{tbl:fit2}.  For further discussion and a comparison with other cluster studies, see Section~\ref{sec:compare}. 

\begin{table}
\begin{minipage}[!ht]{\linewidth}
\caption{Fit coefficients for $\avg{\textup{SFR}}$ and $\avg{\textup{SSFR}}$ of galaxies with the functional form $y=\beta \textup{e}^{\alpha t}$, where t is cosmic time in Gyr.  For the field, we fit both the entire range and allow for a break at $z\sim0.8$.  The reduced $\chi^2$ values for each fit are shown in the last column.}
\label{tbl:fit}
\begin{tabular}{lccc}
\hline
 Coefficients & $\beta^a$  &  $\alpha^b$ & $\chi^2$ \\
\hline
\multicolumn{4}{c}{$y=\avg{\textup{SFR}}$} \\
Clusters ($r<0.5\,$Mpc) & 810$\pm$400 & -0.66$\pm$0.08 & 1.1 \\
Clusters ($r<1\,$Mpc) & 1540$\pm$1100 & -0.76$\pm$0.10 & 1.0 \\
Field & 267$\pm$9 & -0.42$\pm$0.005 & 14.0 \\
\hspace*{5mm}Field ($z<$0.8) & 630$\pm$60 & -0.53$\pm$0.01 & 3.2 \\
\hspace*{5mm}Field ($z>$0.8) & 124$\pm$10 & -0.28$\pm$0.01 & 1.1\\
\hline
\multicolumn{4}{c}{$y=\avg{\textup{SSFR}}$} \\
Clusters ($r<0.5\,$Mpc) & 11$\pm$6 & -0.59$\pm$0.08 & 0.9 \\ 
Clusters ($r<1\,$Mpc) &  15$\pm$12 & -0.66$\pm$0.1& 2.1 \\
Field & 6.4$\pm$0.2 & -0.45$\pm$0.005 & 16.0 \\
\hspace*{5mm}Field ($z<$0.8) & 17$\pm$2 & -0.56$\pm$0.01 & 3.5 \\
\hspace*{5mm}Field ($z>$0.8) & 2.8$\pm$0.2 & -0.30$\pm$0.01 & 1.8 \\
\hline
\end{tabular}
\\
\tiny{$^a$$\beta$ has units of $\Msun$ yr$^{-1}$ for $y=\avg{\textup{SFR}}$ and Gyr$^{-1}$ for $y=\avg{\textup{SSFR}}$.} \\
\tiny{$^b\alpha$ has units of Gyr$^{-1}$. }
\end{minipage}
\end{table}

\FloatBarrier

\subsubsection{Verification at 70$\mu$m}

To verify our procedure of choosing a single SED template to measure L$_{\rm IR}$ and probe the importance of galaxies with unusually warm dust, we have constructed stacks of our galaxy samples at $70\mu$m.  By measuring the L$_{\rm IR}$ of our galaxy samples using one template, we have made two assumptions: 1) that the SEDs of the galaxies in our samples, in particular their dust properties, do not vary significantly over the redshift range we probe, and 2) that the dust properties of our cluster galaxies do not differ systematically from those of our field galaxies.  We outlined some of our justifications for these assumptions in Section~\ref{sec:sfr} and here we further test them by repeating our stacking analysis at 70$\mu$m, a waveband which probes the warm dust component of a galaxy's SED (see Section~\ref{sec:mipsstacking} for the 70$\mu$m stacking procedure).

Figure~\ref{fig:70} shows the ratio of the L$_{\rm IR}$ as derived from the average 70$\mu$m and 250$\mu$m fluxes as a function of redshift for cluster galaxies (red) and field galaxies (purple).  The red shaded region shows the scatter associated with the \citep{kir12} SED template, which was derived from a field galaxy population.   The 70$\mu$m data slightly overpredicts the L$_{\rm IR}$ as compared to 250$\mu$m at $z>0.8$, which may be due to increased warm dust caused by AGN activity, which is known to increase with redshift \citep[e.g.,][]{ued03, ric06, gal09, mar13}.  We verified that the removal of AGN identified in our X-ray and mid-IR data did not significantly change the 70$\mu$m stacked flux measurements.  However, given the uncertainties in the 70$\mu$m fluxes and our AGN selection, this does not necessarily rule out the contributions from lower luminosity AGN. At low redshift, the 70$\mu$m data slightly underpredicts the L$_{\rm IR}$ relative to 250$\mu$m, which may indicate that our chosen template has insufficient cold dust to represent the average low redshift galaxy at the low IR luminosities we are probing (L$_{\rm IR}\sim10^{10}\Lsun$) \citep[e.g.,][]{hwa10, sym13}.  All points, however,  fall within the expected scatter of the SED template, for both cluster and field samples.  This indicates that our use of one SED template to compare cluster and field galaxies as a function of redshift is robust.   When the cluster and field galaxy L$_{\rm IR}^{70\mu m}$ are used to calculate SFRs and SSFRs as a function of cosmic time as in Figure~\ref{fig:masslimited05}, we find that the general trends are preserved, with cluster galaxies in the cluster cores showing a rapid evolution relative to the field.  

\subsection{Evolution of Cluster and Field Galaxies with Respect to Stellar Mass}
\label{sec:mass}

\begin{figure*}
\begin{minipage}[b]{0.45\linewidth}
\centering
\includegraphics[width=1.1\linewidth, trim=5mm 0 0 0, clip]{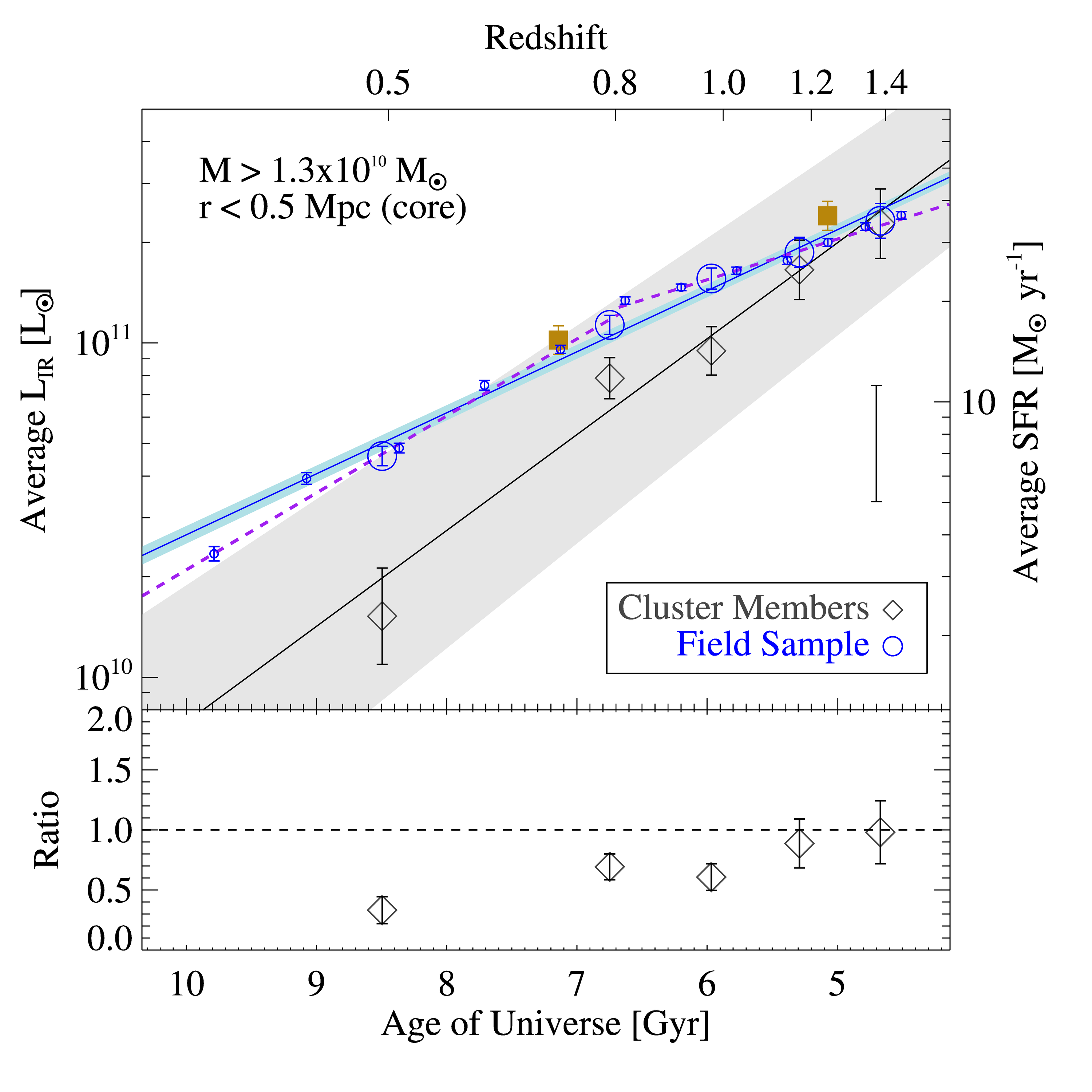}
\end{minipage}
\hspace{0.5cm}
\begin{minipage}[b]{0.45\linewidth}
\centering
\includegraphics[width=1.1\linewidth, trim=5mm 0 0 0, clip]{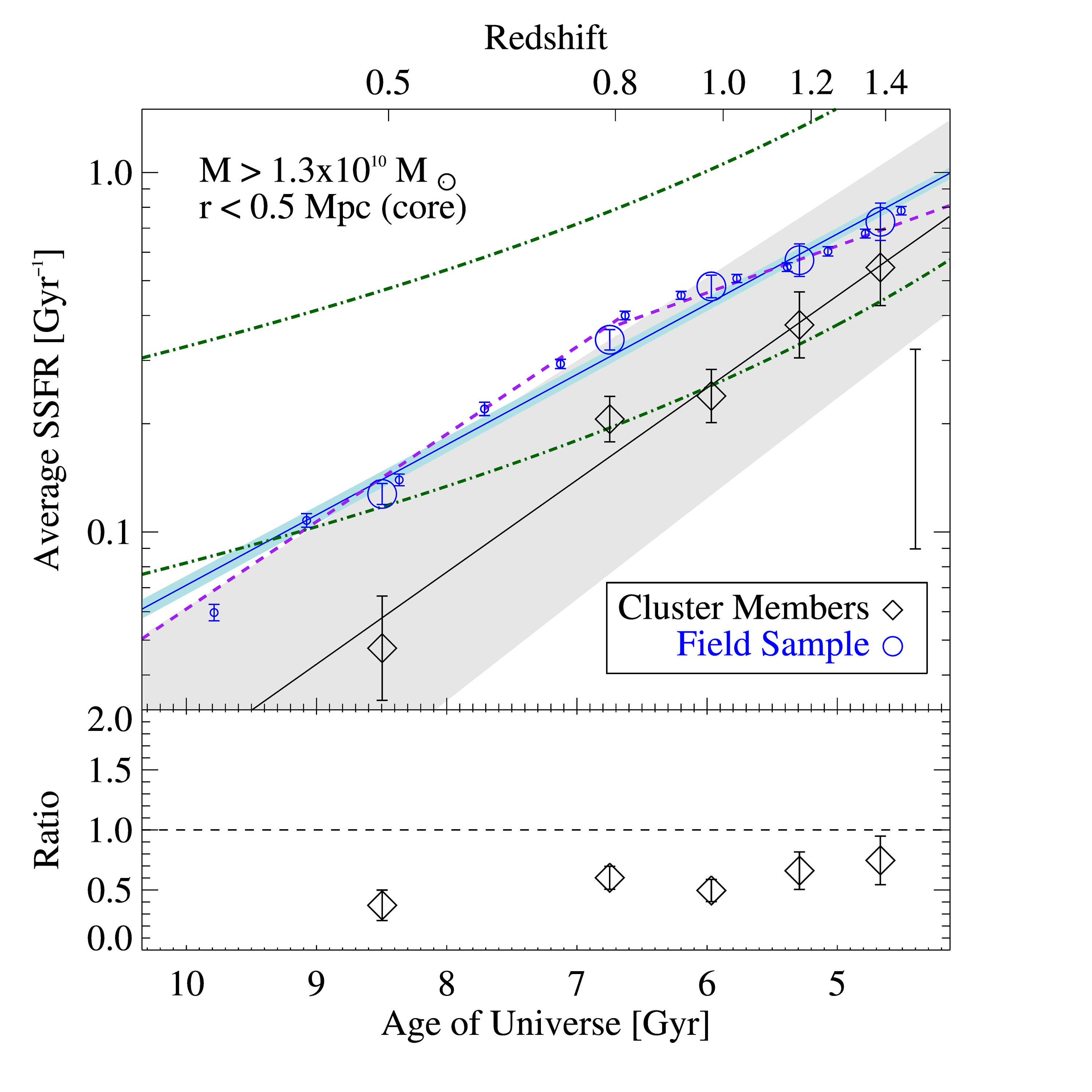}
\end{minipage}
\caption{Evolution of the average L$_{\rm IR}$, SFR, and SSFR in cluster galaxies relative to the field.  (Left) The top panel shows the $\avg{\mathrm{L_{IR}}}$ and $\avg{\textup{SFR}}$ of cluster galaxies (black diamonds) within a projected radius of 0.5~Mpc versus field galaxies (blue circles) as a function of redshift, while the bottom panel shows the ratio of $\avg{\mathrm{L_{IR}}}$ for cluster to field galaxies.  The large blue circles are the field stacked in the same redshift bins as cluster galaxies while the smaller blue circles are field galaxies in higher resolution redshift bins with width 0.1.  We fit the cluster member and high resolution field galaxy stacks with the function $y=\beta \textup{e}^{\alpha t}$ (black and blue solid lines respectively) to quantify the rapid rise of the SF activity in cluster members as a function of redshift as compared to the field.  The shaded regions show the $1\sigma$ errors on the fits.  The dashed purple line shows that the high resolution field stacks are better fit by two $y=\beta \textup{e}^{\alpha t}$ functions, broken at z=0.8.  The filled yellow squares are field galaxies from UDS, stacked using S\lowercase{{\sc imstack}} \citep{vie13}. (Right) The same for $\avg{\textup{SSFR}}$, which shows that the differences in average SF properties between cluster and field galaxies cannot be fully accounted for with mass differences between the two populations.  The green, dashed-dot lines denote the boundaries of the infrared Main Sequence  as defined in \citet{elb11}.  The large error bar represents the uncertainties associated with the \citet{kir12} template SED and stellar mass estimates.}
\label{fig:masslimited05}
\end{figure*}

We examine the average L$_{\rm IR}$, SFR, and SSFR  as a function of stellar mass by breaking our cluster and field samples into two stellar mass bins: 1.3$\e{10} < $ M$_{\star}< 6.3\e{10} \Msun$ and M$_{\star} > 6.3\e{10} \Msun$, chosen as roughly the middle value in the mass range we probe.  The results are as follows.  In the cluster cores, we find that the $\avg{\textup{SSFR}}$ of the higher mass galaxies (Figure~\ref{fig:mass}, upper right) is suppressed at $\sim70\%$ of the field SSFR but otherwise shows no strong differential evolution with respect to the field as a function of redshift.  Conversely, in the cores+outskirts, the higher mass galaxies show a stronger evolution relative to the field galaxies (lower right).  This suggests that multiple mechanisms may be responsible for the evolution of high mass galaxies in the cores versus the outskirts.  The lower mass cluster galaxies (left), on the other hand, are the primary drivers of the field-like star formation activity in the full galaxy population at high redshift (Figure~\ref{fig:masslimited05}).  This is true in both the cores (upper left), where the lower mass galaxies show field-like star formation in the $\avg{z}=1.2-1.4$ bins, and in the core+outskirts (lower left), where the low mass galaxies are experience enhanced star formation above the field level.  The average L$_{\rm IR}$ and SFRs in these stellar mass bins show the same trends as the  $\avg{\textup{SSFR}}$.

\begin{table}
\begin{minipage}[!ht]{0.9\linewidth}
\caption{Fit coefficients for $\avg{\textup{SFR}}$ and $\avg{\textup{SSFR}}$ of the functional form $y\sim(1+z)^n$.   The reduced $\chi^2$ values for each fit are shown in the last column.}
\label{tbl:fit2}
\begin{tabular}{lcc}
\hline
 Coefficients & $n$  & $\chi^2$ \\
\hline
\multicolumn{3}{c}{$y=\avg{\textup{SFR}}$} \\
Clusters ($r<0.5\,$Mpc) &  $5.6\pm0.6$ & 1.8 \\
Clusters ($r<1\,$Mpc) &  $5.9\pm1.0$ & 0.6 \\
Field &  $3.9\pm0.4$ & 39.0 \\
\hline
\multicolumn{3}{c}{$y=\avg{\textup{SSFR}}$} \\
Clusters ($r<0.5\,$Mpc) & $5.0\pm0.7$ & 1.33 \\ 
Clusters ($r<1\,$Mpc) &  $5.3\pm1.1$ & 1.6 \\
Field & 4.0$\pm$0.05 & 32.5 \\
\hline
\end{tabular}
\end{minipage}
\end{table}

\FloatBarrier

%\emph{\color{red}In order to compare with other studies (Bai+09, Popesso+12, Webb+13), I also calculate the mass-normalized integrated SFR, $\Sigma$SFR/M$_{halo}$, for $r< 0.5$ Mpc.  To do this, I assume that all the of the clusters have M$_{halo}$=10$^{14} \Msun$ and, if this assumption is at all appropriate (do we have an idea of the expected scatter and is it redshift dependent?), then I find $\Sigma$SFR/M$_{halo}=(14.2\pm6.7)(1+z)^{4.2\pm0.7}$ or $\Sigma$SFR/M$_{halo}\propto (z)^{1.96\pm0.4}$. Bai+09 and Web+13 find $\Sigma$SFR/M$_{halo}\propto (1+z)^{5.3\pm1.2}$ and $\Sigma$SFR/M$_{halo}\propto (1+z)^{5.4\pm1.9}$, respectively.  The Bai+09 sample is $\sim$10$^{15}\Msun$ clusters down to 2$\Msun$ yr$^{-1}$ and out to 0.5$r_{200}$ for $z$=0-0.8.  Webb+13 looked at 10$^{14}$-10$^{15}\Msun$ clusters from $z$=0.3-1 with limiting LIR depth$\sim$10$^{11}\Lsun$ (no stellar mass limit).  Popesso+12 fit $\Sigma$SFR/M$_{halo} \propto (z)^{1.77\pm0.4}$ for $>$3$\e{14}$ clusters out to $z$=0.85 and $\Sigma$SFR/M$_{halo} \propto (z)^{1.33\pm0.3}$ for $>$1$\e{14}$ clusters, I find $\Sigma$SFR/M$_{halo}\propto (z)^{1.96\pm0.4}$.}

%\emph{\color{red}Things look consistent to first order and I will do a more careful comparison later unless there is a reason for us not to look at this parameter because the cluster masses are too unconstrained.  If we decide to examine this parameter, I will include a plot showing our results compared to these other studies.}

\newpage

\begin{figure*}
\begin{minipage}[b]{0.45\linewidth}
\centering
\includegraphics[width=1.1\linewidth, trim=5mm 0 0 0, clip]{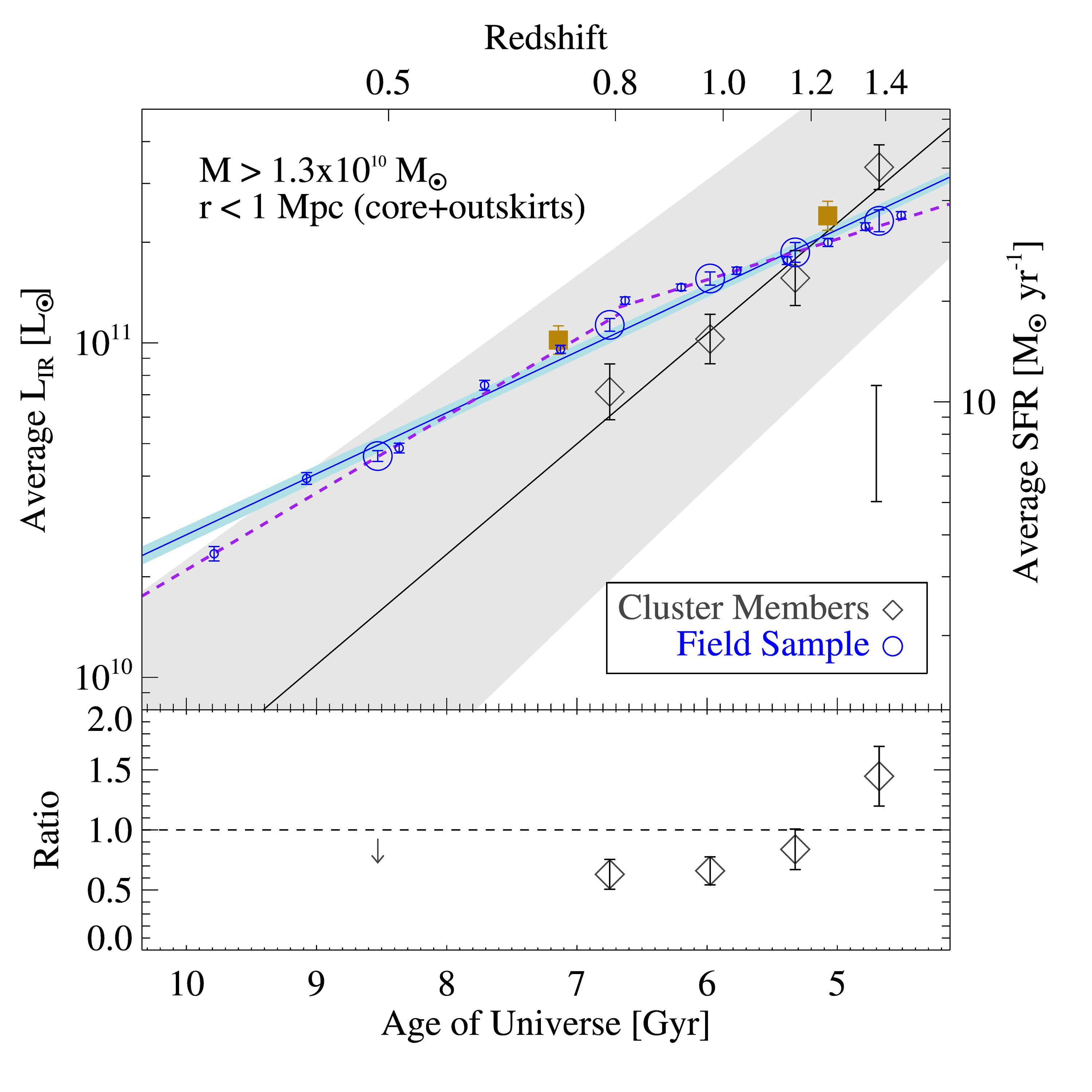}
\end{minipage}
\hspace{0.5cm}
\begin{minipage}[b]{0.45\linewidth}
\centering
\includegraphics[width=1.1\linewidth, trim=5mm 0 0 0, clip]{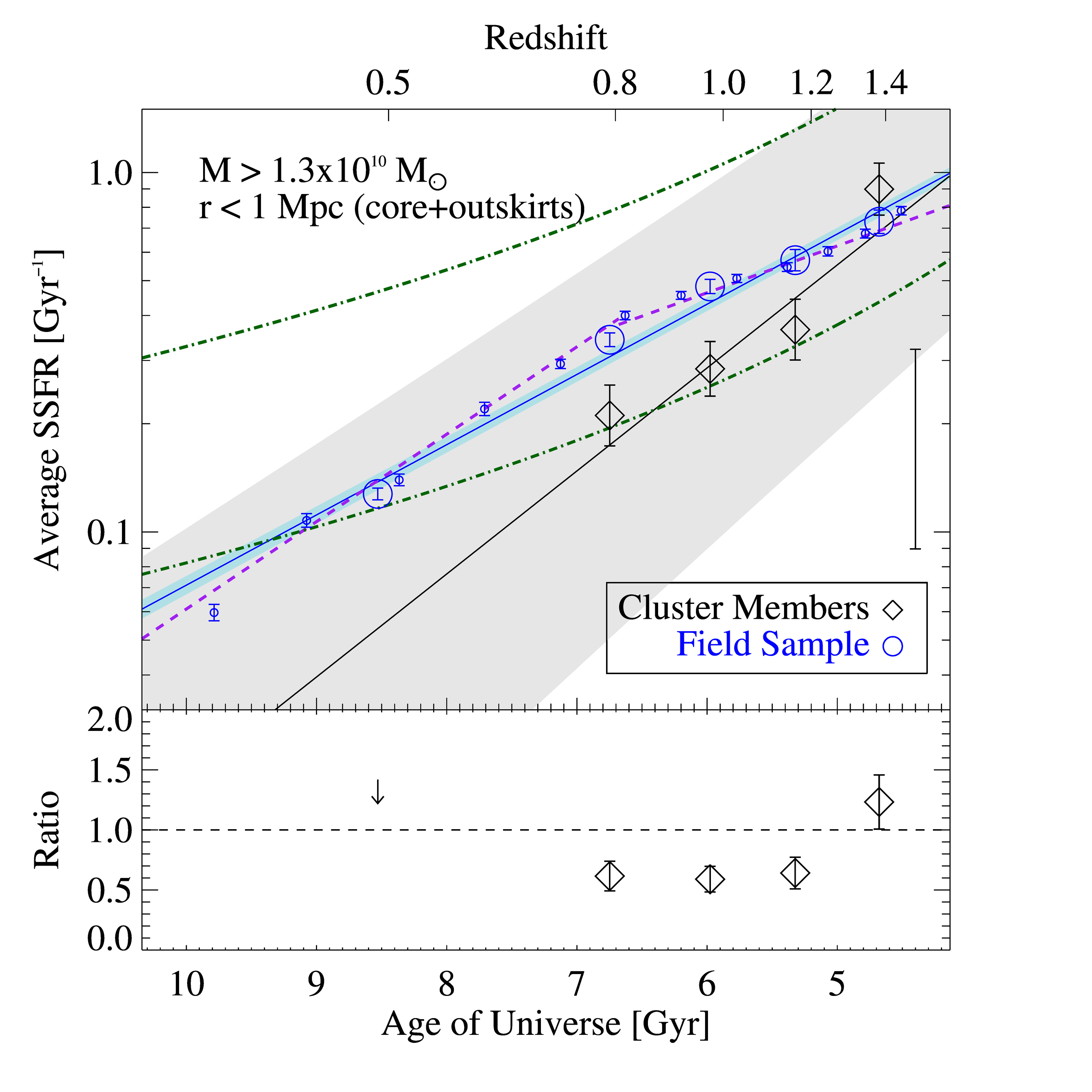}
\end{minipage}
\caption{The same as Figure~\ref{fig:masslimited1}, but with a projected radius of 1~Mpc (cluster core+outskirts).}
\label{fig:masslimited1}
\end{figure*}

\FloatBarrier

\subsection{The Evolution of Star-Forming, Blue Galaxies in Clusters versus the Field}
\label{sec:redsequence}

In this section, we separate out star forming galaxies in order to analyse whether the evolutionary trends we see are due to a change in the properties of currently star forming galaxies.  As part of the process of deriving photometric redshifts, each galaxy is matched to a best fit template chosen to represent late-type galaxies (Sb, Sc, Sd, Spi4, and M82) and early-type galaxies (Ell5, Ell13, S0, and Sa) from \citet{pol07} using optical and near-infrared photometry (see Section~\ref{sec:redshifts}).  Whether a galaxy is best-fit to a late-type or early-type template depends predominantly on the strength of its 4000\AA break.  This allows us to roughly separate our galaxy samples into ``blue" (late-type) and ``red" (early-type) sub-samples.  This selection is similar to traditional methods of using rest-frame optical colors which bracket the 4000\AA$\,\,$ break to separate galaxies into star forming and quiescent categories.  The process of matching the best-fit template for deriving photometric redshifts is applied in the same way to both cluster and field galaxies, meaning that we can consistently compare blue or red galaxies in the cluster to those in the field using this selection.

 We note that, using this selection technique, galaxies fit to early type templates may be truly passive or may be star forming galaxies that are so heavily dust-obscured as to look red.  By matching to MIPS 24$\mu$m, we find that 15-30$\%$ of galaxies best-fit by early-type templates have a corresponding MIPS detection within 4$^{\prime\prime}$.  Unfortunately, the MIPS catalogue is too shallow to detect the characteristic L$_{\rm IR}$ of our sample at z$\gtrsim$1 and so gives an incomplete census of contamination as well as introducing complications from AGN contamination.    As such, we focus on the blue galaxies as a representative sample of star forming galaxies with non-extreme dust properties and determine their average L$_{\rm IR}$, SFR, and SSFR properties, with the caveat that we are likely missing some fraction of heavily dust-obscured star formation, a fraction which will grow more significant with increasing redshift. 

In Figure~\ref{fig:redsequence}, we compare the average SFR (left) and SSFR (right) of blue galaxies in the cluster cores versus the field.  We find that the evolution of $\avg{\textup{SFR}}$ shows an increase with redshift compared to the field, as we saw with the full sample in Figure~\ref{fig:masslimited05}; however, when the stellar mass of the blue galaxies is taken into account for the $\avg{\textup{SSFR}}$, the star forming galaxies no longer show a strong evolution relative to the field over time, though they are suppressed at $\sim70\%$ of the field SSFR.  At $\avg{z}$=1.4, the average SFR in the cluster cores is consistent with the field, but the average SSFR is lower.  This may be an indication that the stellar mass function of blue, star forming galaxies is different in cluster versus the field at these redshifts.  At lower redshifts, on the other hand, the average SFR and SSFR are both quenched below the field level. Taken together, these two plots indicate that both the SFRs and stellar mass distributions in cluster galaxies relative to the field may be different over our redshift range.   In the core+outskirts (not shown), the average SFR and SSFR behave in the same manner with the exception of the $\avg{z}$=1.4 bin, which again has enhanced star formation of 1.7 times the field SFR and 1.2 times the field SSFR.  

We compare our results to a recent study which looked at the average SSFRs in star forming cluster galaxies from $z=0.15-0.3$.  \citet{hai13} too found that the SSFR does not show a strong differential evolution relative to the field, but that the average SSFR is suppressed below the field level.  We show the \citet{hai13} results in Figure~\ref{fig:redsequence} (right), where we also indicate the region which corresponds to the infrared Main Sequence \citep{elb11}.  Our star-forming galaxy samples, both cluster and field, fall on the Main Sequence at all redshifts.

\section{Discussion}
\label{sec:disc}

As cluster studies push to higher and higher redshifts, the challenge becomes not just to explain the signature properties of local clusters -- the strong, red sequence of passively evolving galaxies -- but to constrain the epoch in which clusters were engaging in active mass build-up, with the star formation necessary to assembly present-day massive ellipticals.  Using a uniform sample of clusters $(\sim10^{14} \Msun$), we have demonstrated that the average 250$\mu$m flux (and by extension the dust-obscured SFR) of cluster galaxies is quenched below the field level across most of cosmic time, $\sim8$ Gyr, but with a rapid evolution in which the average SFR of cluster galaxies draws even with the field in the cluster cores at $z\gtrsim$1.2, with enhanced SF above the field level in the cluster outskirts.   We measure an e-folding time for the evolution in the cluster cores of $\sim1.5$ Gyr over $0.3<z<1.5$.  This is consistent with the findings of \citet{bro13}, who looked at cluster members detected at 24$\mu$m from $1.0<z<1.5$ and found a sharp transition from active to quenched SF.  Here we explore what mechanisms might be responsible for the evolution we observe.

\begin{figure}
\centering
\includegraphics[width=1\linewidth, trim=80mm 0 0 0, clip]{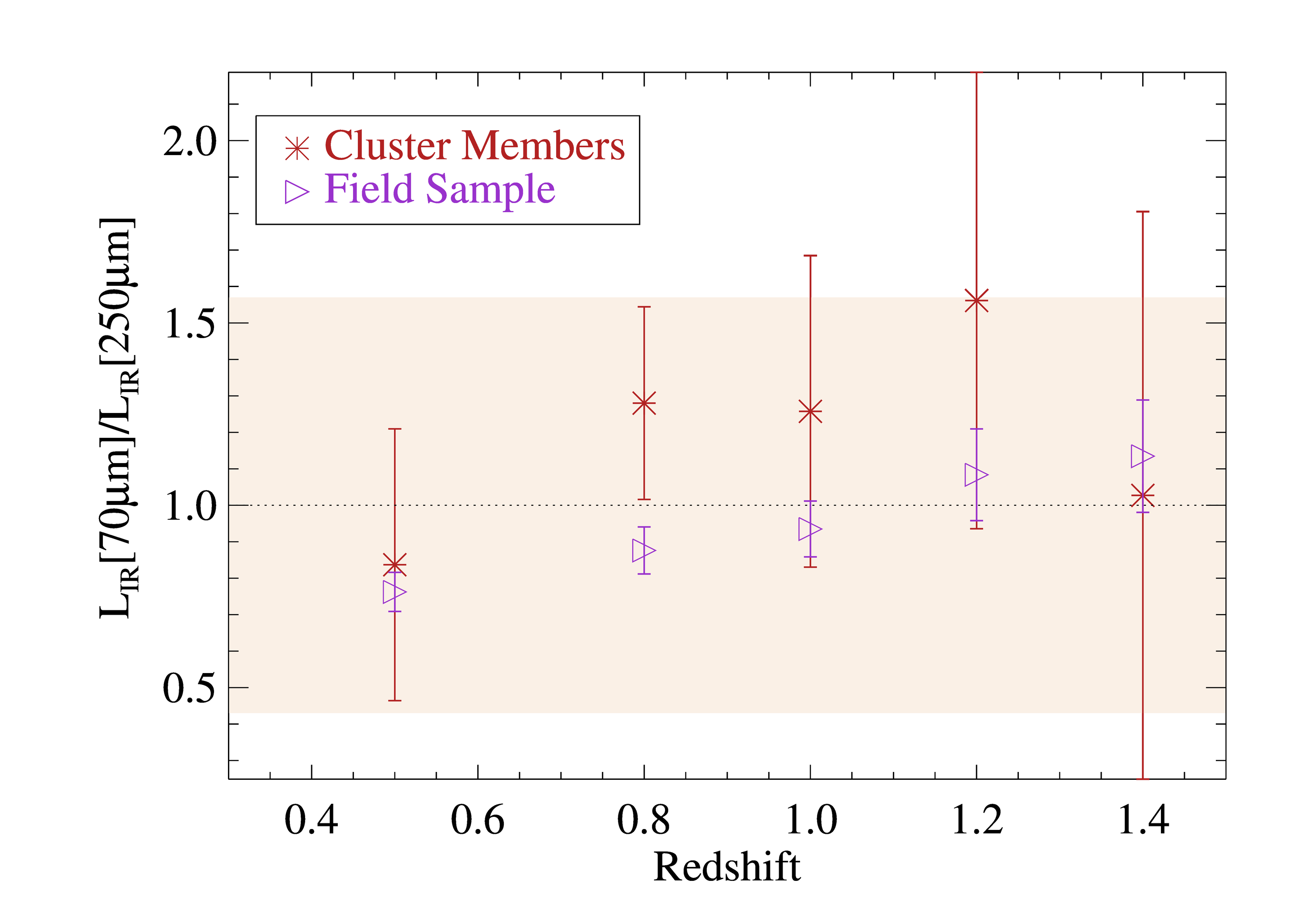}
\caption{The ratio of the L$_{\rm IR}$ for cluster (red stars) and field galaxies (purple triangles) derived from stacking at 70$\mu$m and 250$\mu$m.  The red, shaded band shows the scatter associated with the \citet{kir12} SED template used to calculate the L$_{\rm IR}$.  All points fall within the expected scatter of the SED template, indicating that the template represents both the average warm and cold dust properties of the cluster and field galaxy samples as a function of redshift.}
\label{fig:70}
\end{figure}

\FloatBarrier

\begin{figure*}
\begin{minipage}[b]{\linewidth}
\centering
\includegraphics[width=0.9\linewidth, trim=5mm 0 0 0, clip]{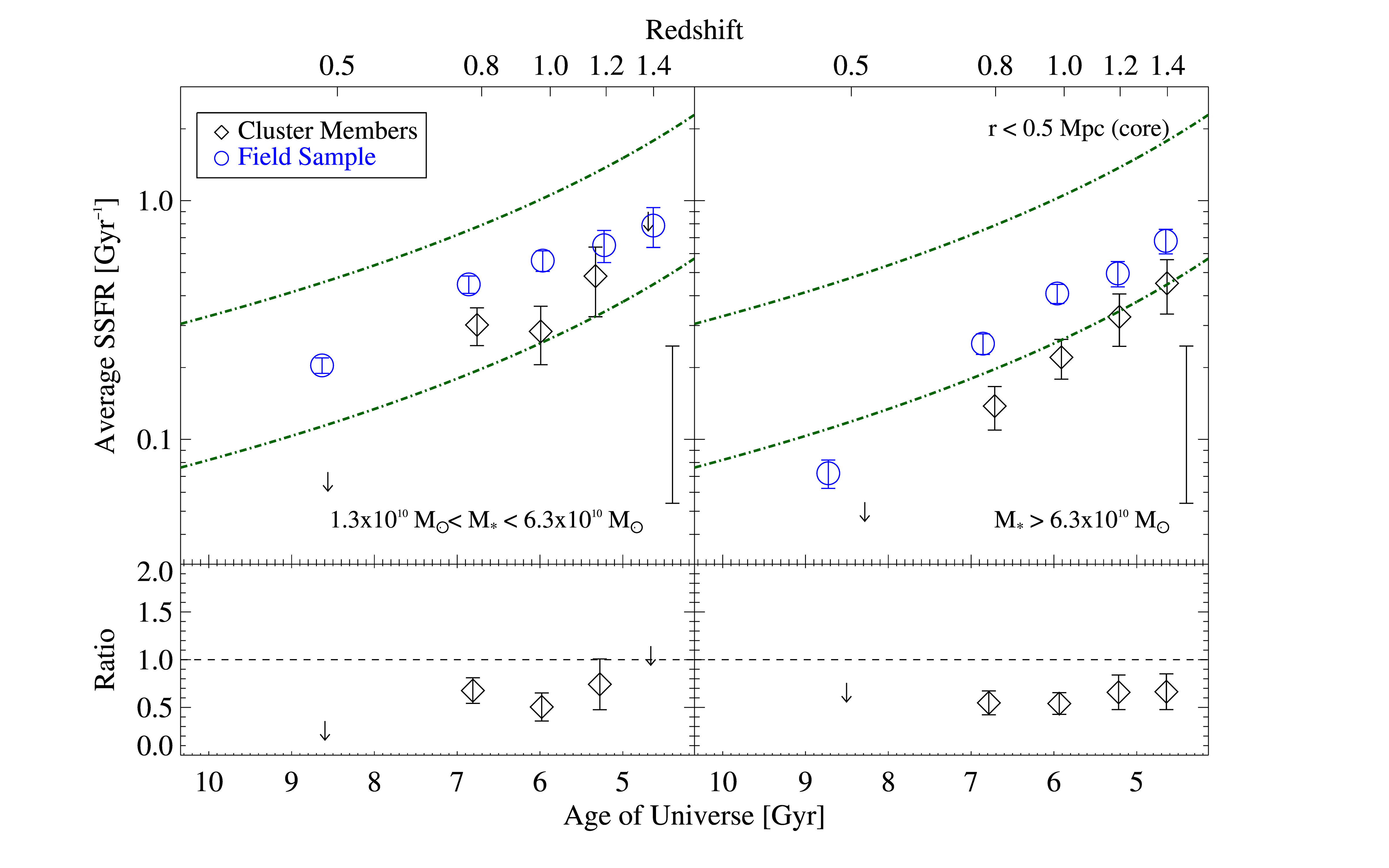}
\end{minipage}
\hspace{0.5cm}
\begin{minipage}[b]{\linewidth}
\centering
\includegraphics[width=0.9\linewidth, trim=5mm 0 0 0, clip]{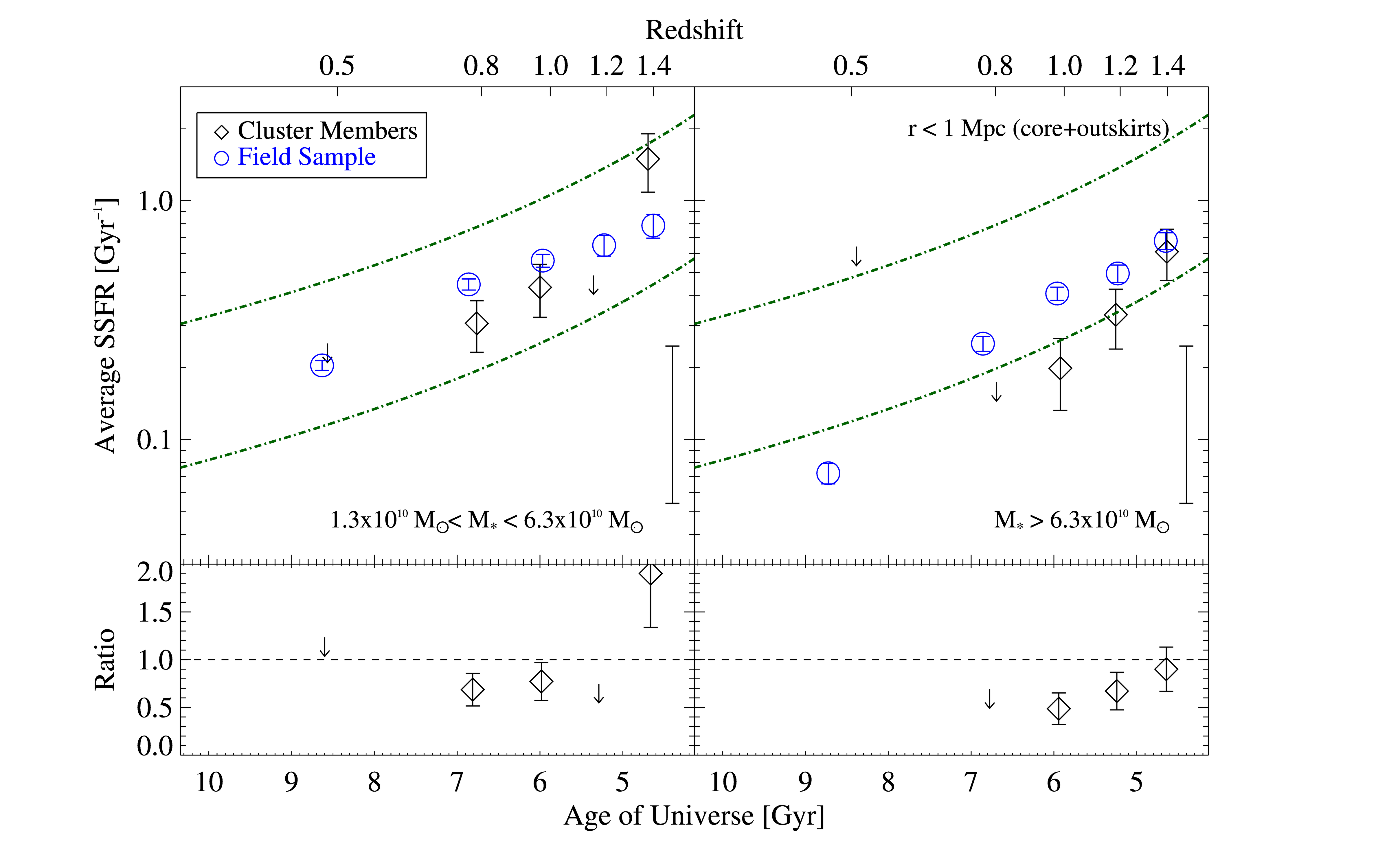}
\end{minipage}
\caption{The $\avg{\textup{SSFR}}$ for cluster (black diamonds) and field (blue circles) galaxies as in Figure~\ref{fig:masslimited05}, but for mass bins $1.3\e{10}<$ M$_{\star}<6.3\e{10}\Msun$ (left) and M$_{\star}>6.3\e{10}\Msun$ (right).  The top panels show cluster members out to projected radius of 0.5~Mpc (core) and the bottom panels show out to 1~Mpc (core+outskirts).  In the cluster cores, the evolution in star formation activity seems to be dominated by the lower mass galaxies, as the higher mass galaxies show no strong differential evolution with respect to the field, though they are suppressed below the field level at all redshifts.  When the outskirts are included we see that all cluster galaxies are on average evolving more rapidly than the field, with the lower mass galaxies showing enhancement over the field in the highest redshift bin. The green, dashed-dot lines denote the boundaries of the infrared Main Sequence  as defined in \citet{elb11}. }
\label{fig:mass}
\end{figure*}

\FloatBarrier

\begin{figure*}
\begin{minipage}[b]{0.45\linewidth}
\centering
\includegraphics[width=1.1\linewidth, trim=5mm 0 0 0, clip]{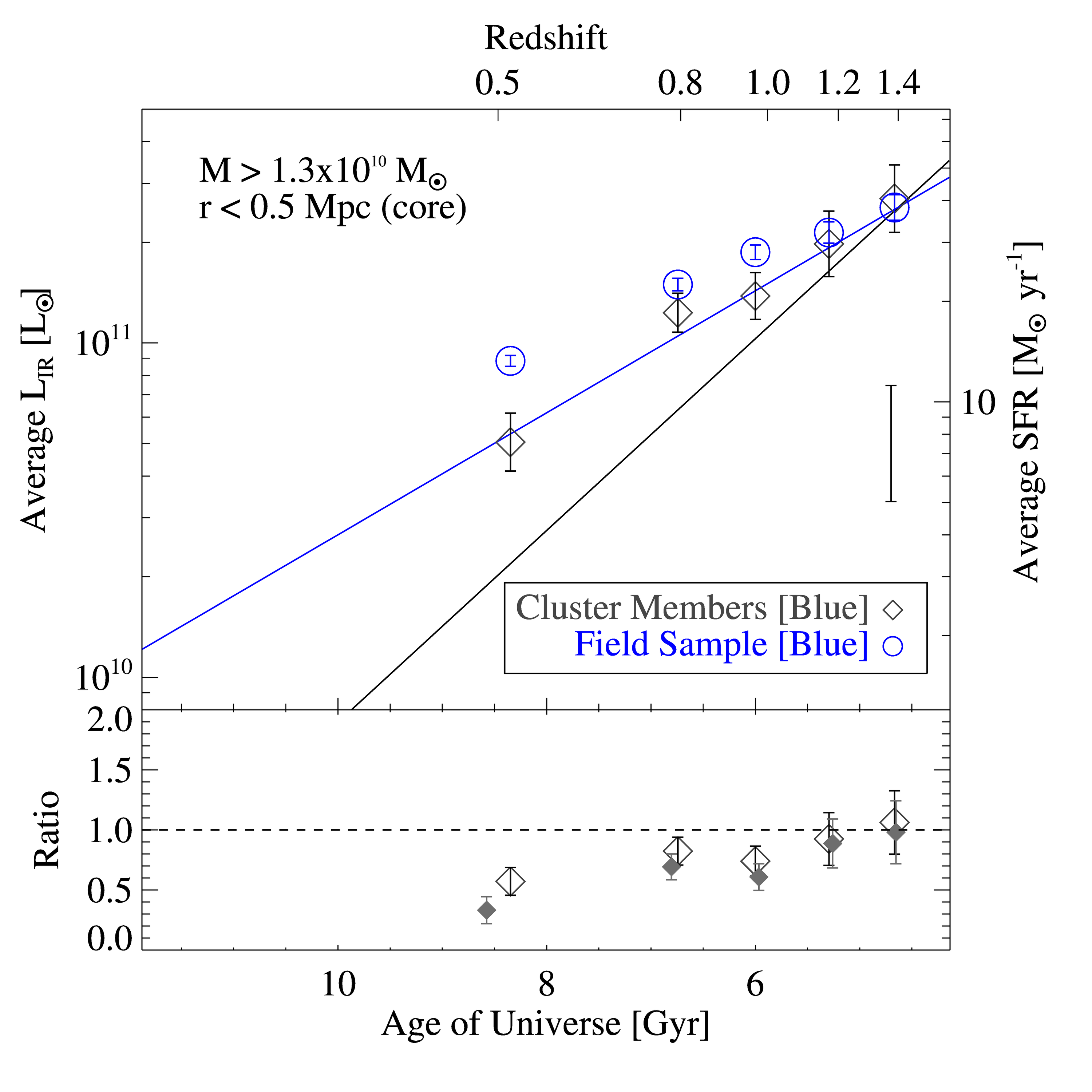}
\end{minipage}
\hspace{0.5cm}
\begin{minipage}[b]{0.45\linewidth}
\centering
\includegraphics[width=1.1\linewidth, trim=5mm 0 0 0, clip]{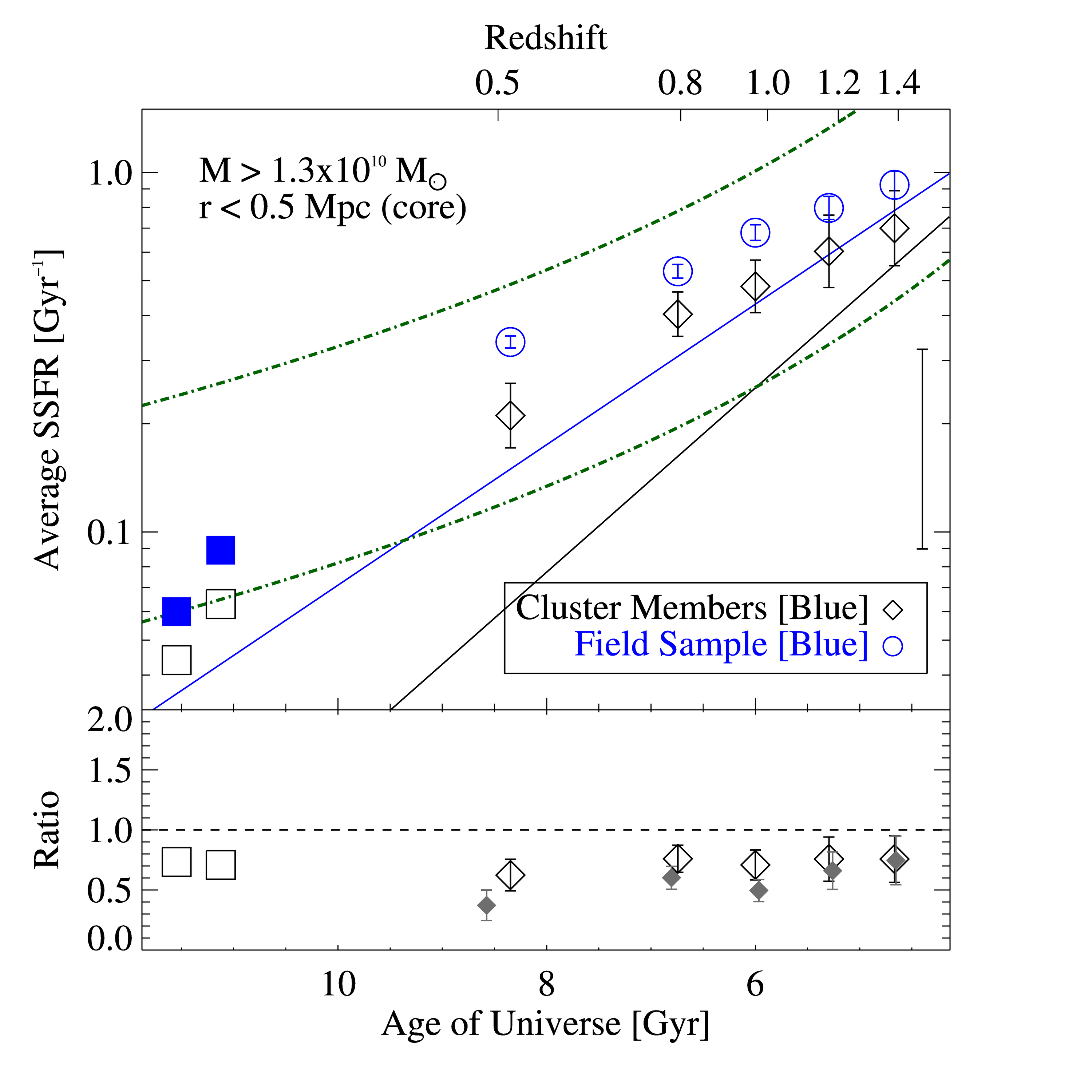}
\end{minipage}
\caption{The evolution of star forming galaxies in cluster cores versus the field.  The average L$_{\rm IR}$ and SFR (left) and average SSFR (right) for cluster galaxies (black diamonds) versus field galaxies (blue circles) as in Figure~\ref{fig:masslimited05} for blue galaxies only.  Though the average SFR shows a rapid decline with cosmic time compared to the field, the average SSFRs show no strong differential evolution with respect to the field.  The blue and black lines show the evolution of all galaxies with cosmic time, as seen in Figure~\ref{fig:masslimited05}.  In the bottom panels, the filled, gray diamonds show the ratio of L$_{\rm IR}$ for all cluster to field galaxies as in Figure~\ref{fig:masslimited05}.  The dashed-dot green lines show the region of SSFR as a function of redshift denoted the infrared Main Sequence \citep{elb11}.  The blue and black squares are average SSFRs for star forming cluster (open squares) and field (filled squares) galaxies at $z=0.18-0.22$ \citep{hai13}.}
\label{fig:redsequence}
\end{figure*}



%A recent study of a range of environments in the COSMOS field also found an abrupt ($\sim1$ Gyr) increase in the fraction of passive galaxies in dense regions at z$\sim1.2$ \citep{sco13}.  Though that study did not probe the characteristic densities of clusters, this evolution is also consistent with several studies which found increasing star formation and/or an increase in the star forming fraction toward dense regions in individual high redshift ($z>1.4$) clusters \citep[e.g.,][]{tra10, hil10, hay11, fas11, tad12}.  These studies, however, suffer from the effects of cluster-to-cluster variations and often rely on optical emission line star formation tracers in an era when dust-obscured star formation is dominating the SF budget.  

\subsection{Quenching Mechanisms}
\label{sec:quenching}

Several pieces of evidence presented here give us clues about the processes involved in the quenching of star formation activity in cluster galaxies.  We find that in the cluster cores ($r<0.5\,$Mpc), the full population of cluster galaxies (Figure~\ref{fig:masslimited05}) shows significant quenching over the redshifts we probe, starting with field-like SF activity at $z\gtrsim1.2$ and quenching with an e-folding time of $\sim1.5\,$Gyr.  This is considerably faster than the e-folding time of SF in field galaxies, $\sim$2.4 Gyr, where galaxy evolution is likely driven by mass-quenching, gas accretion, and/or AGN \citep{mo10}.  This rapid evolution is seen in both the average SFR and SSFR, the latter suggests that these trends cannot be fully explained by a different stellar mass functions for cluster and field galaxies. 
%The lack of enhanced SF activity in cluster galaxies relative to the field (with the exception of the cluster outskirts at $\avg{z}=1.4$) indicates that processes which cause the rapid consumption of gas through starbursts (e.g. mergers) are not the dominant mechanism for quenching.}

When broken into sub-populations, our cluster galaxies suggest that multiple processes are likely operating in these clusters.  High mass cluster galaxies (M$\,>6.3\e{10}$) in the cores show no strong evolution relative to the field, which may indicate that their evolution is dominated by mass-quenching.  This is consistent with the results of \citet{pen10}, who found that galaxies of these masses are dominated by internal evolution regardless of environment.  High mass galaxies in the cluster outskirts, however, do show a more rapid evolution relative to the field. Lower mass galaxies show a more rapid evolution at all redshifts and radii, with field-like star formation in the cores at high redshift and enhanced star formation in the outskirts. 

 Blue, star forming galaxies show a strong evolution relative to the field in their SFRs, but no strong evolution in their SSFRs.  Unlike the full galaxy populations, this suggests that the evolution in blue galaxy SFRs could be fully explained by different stellar mass functions between cluster and field for blue galaxies specifically.  This would be consistent with studies of low redshift massive clusters, where measures of the H$\alpha$ luminosity function \citep{kod04} and mid-IR SFRs \citep{bai09, hai09} were found to be largely independent of environment.  \citet{hai13} found a similar trend with the SSFR in low redshift clusters (see Section~\ref{sec:compare}).

Taken together, these observations suggest that multiple cluster-specific processes may be driving the evolution of sub-populations of cluster galaxies in different cluster regions, while other dusty galaxies (high mass, core galaxies) may be dominated by mass-quenching.  If the trends seen in the SFRs of blue, star forming galaxies can be explained as differences in the stellar mass distribution of cluster galaxies, then the evolution of the full population may be driven by the rapid transition of star forming galaxies to the quiescent galaxy population through the effective shut down of SF.  This is supported by \citet{bro13}, who found a strong transition to lower SFRs below $z\sim1.3-1.4$ in $z>1$ ISCS clusters using MIPS 24$\mu$m observations and concluded that these trends can be explained by merger-driven star formation followed by rapid AGN quenching in $z\gtrsim1.5$ clusters.  These observations further support \citet{muz12}, who found a lack of correlation between SSFR and D$_n$(4000) in star forming galaxies with environment at $z\sim1$ and a high post-starburst fraction. They concluded that star forming galaxies are transitioning to the quiescent population on rapid time-scales at higher redshifts.  This transition would require that the cold gas which fuels star formation in galaxies be consumed, heated, or removed.  In this work, we have observed evidence for the previously suggested mergers at high redshifts in the cluster outskirts; however, we do not see enhanced star formation on average at lower redshifts and radii (though this does not rule out dry mergers).  A more likely scenario for ongoing quenching at lower redshifts and in the cluster cores may involve the removal of gas.  This is supported by local observations, which have found cluster galaxies to be increasingly deficient in HI gas close to cluster centers  \citep{hay84, sol01, hug09} as well as cluster galaxies with truncated gaseous disks \citep[e.g.,][]{koo04,koo06} and long extra-galactic tails of HI gas \citep{chu07}.   The two main processes that remove gas in galaxies in dense environments are strangulation \citep{lar80} and ram pressure stripping \citep{gun72}.  For a review of cluster processes in general, see \citet{bos06}.

 Strangulation, the removal of loosely-bound hot halo due to the ICM and global tidal field of the clusters,  is capable preventing the re-fueling of galaxies over several Gyr.    %This timescale is too long to be responsible for the rapid evolution we see in the full cluster galaxy population, however, it may play a role in the lower $\avg{\textup{SSFR}}$ we see for blue and high mass cluster galaxies.  
Unlike their analogues in the field, cluster galaxies can no longer accrete fresh, cold gas once they enter a region with a hot, dense ICM.  This lack of fresh gas may lower their SFR relative to field galaxies on long time-scales and we suggest this may be responsible for the lower SSFRs of high mass galaxies in the cluster cores.
%, while their evolution is still be dominated by internal quenching as they consume the gas that remains.  
%Our high mass galaxy bin, M$_{\star}>6.3\e{10}\Msun$, is firmly in the mass-quenching regime of \citet{pen10}.  Our blue galaxy sample exhibits the same behavior, showing {\bf a systematic suppression of star formation, but no strong evolution with respect to field galaxies as a function of cosmic time.}   This suggests that the long term quenching of our blue galaxy sample is due to the strangulation of gas lowering their SFRs and that their overall evolution is dominated by mass-quenching.  Mass-quenching will continue until the galaxy's gas is consumed or some other process strips it out.

Ram pressure stripping (RPS), the removal of the ISM by the hot ($\sim10^7-10^8$ K), dense ($\sim10^{-3}-10^{-4}$ atoms cm$^{-3}$) ICM, can operate efficiently on galaxies with high orbital velocities ($\gtrsim1000$ km s$^{-1}$), loosely bound ISMs such as in intermediate to low mass galaxies, and in clusters with short crossing times.  Hydrodynamical simulations of individual galaxies using the \citet{gun72} RPS estimation found the timescale for gas removal to be $\sim10-100$ Myr \citep{aba99, qui00, mar03, roe06, roe07, kro08}.  As such, lower mass galaxies near the cluster cores may see their gas stripped away on short time-scales, stopping their SF and adding them to the passively evolving galaxy fraction.

\subsubsection{A Back-of-the-envelope Calculation for Gas Depletion}
\label{sec:gasdepletion}

By making some simplifying assumptions, we can link our measured $\avg{\mathrm{L_{IR}}}$ for cluster and field galaxies to the fraction of galaxies which retain gas between $z=1$ and $z=0.5$.  We first assume that if a galaxy has gas, then it contributes a fixed amount to the average L$_{\rm IR}$, $\ell_{\rm IR,g}$; if it contains no gas, it contributes nothing.  If the fraction of galaxies that retain their gas is given by $f_g(z)$  and the total number of galaxies is $N_g(z)$ then 

\begin{equation}\label{eqn:rps1}
\begin{split}
\avg{\mathrm{L_{IR}}(z)} = \frac{\Sigma \mathrm{L_{IR}}(z)}{N_{g}(z)} \approx \frac{f_g(z) N_g(z) \ell_{\mathrm{IR}, g}(z)}{N_g(z)} \\
  = f_g(z)\ell_{\mathrm{IR}, g}(z) .
\end{split}
\end{equation}

Consider the field-normalized ratio of the average L$_{\rm IR}$ of cluster galaxies at $z=1$ to $z=0.5$, $Q$,

\begin{equation}\label{eqn:rps2}
\begin{split}
Q &= \frac{\avg{\mathrm{L_{IR}^{cl}}(z=1)}/\avg{\mathrm{L_{IR}^{cl}}(z=0.5)}}{\avg{\mathrm{L_{IR}^{field}}(z=1)}/\avg{\mathrm{L_{IR}^{field}}(z=0.5)}} \\ 
\\
&= \frac{\left[\dfrac{f_g^{cl}(z=1)}{f_g^{cl}(z=0.5)}\right]  \left[\dfrac{\ell^{cl}_{\mathrm{IR}, g}(z=1)}{\ell^{cl}_{\mathrm{IR}, g}(z=0.5)}\right]} {\left[\dfrac{f_g^{field}(z=1)}{f_g^{field}(z=0.5)}\right]\left[\dfrac{\ell^{field}_{\mathrm{IR}, g}(z=1)}{\ell^{field}_{\mathrm{IR}, g}(z=0.5)}\right]}
\end{split}
\end{equation}
We further assume that the fraction of galaxies with gas in the field does not change significantly, $\frac{f_g^{field}(z=1)}{f_g^{field}(z=0.5)}=1$, and that, in the absence of gas stripping, the contribution to the total IR luminosity for cluster galaxies which retain their gas is equal to contributions from field galaxies: $\ell_{\rm IR, g}^{cl}(z)$ = $\ell_{\rm IR, g}^{field}(z)$ (this assumption breaks down on the time-scales of strangulation).  This simplifies $Q$ to a simple ratio of the fraction of galaxies that retain gas in clusters at $z=1$ to at $z=0.5$: $Q\approx\frac{f_g^{cl}(z=1)}{f_g^{cl}(z=0.5)}$.  From Equation~\ref{eqn:rps2}, the ratio of our average L$_{\rm IR}$ for cluster  and field galaxies across $z=0.5-1$ is then approximately the fraction of galaxies which retain gas over the same redshift range.  We calculate $Q\approx1.8\pm0.7$ from our observations at $r<0.5\,$Mpc (Figure~\ref{fig:masslimited05}). 

\subsubsection{Comparison to a Ram Pressure Stripping Simulation}
\label{sec:rps}

\citet{tec10} performed a self-consistent estimation of the effects of ram pressure stripping in moderate to high mass clusters using a semi-analytic model of galaxy formation combined with hydrodynamical simulations of galaxy clusters.  They calculated the fraction of galaxies which have been stripped of their gas as a function of cluster-centric radius and redshift, finding that out to the virial radius of $\sim10^{14}\,\Msun$ clusters, this fraction increases by a factor of 2 from z$=1$ to $z=0.5$.  Their simulations consider galaxy velocities of 700-3000 km s$^{-1}$ and note that the ICM density increases an order of magnitude from $z=1$ to the present day (with $\rho_{\rm ICM}\sim10^{-6}-10^{-3}$ atoms cm$^{-3}$ at $z=1$).

From \citet{tec10}, we determine the simulated fraction of cluster galaxies that \emph{retain} their gas from z=1 to z=0.5 at a radius of 0.5 Mpc for $\sim10^{14}\,\Msun$ clusters is $Q=1.5\pm0.3$ \citep{tec10}, while our observations show $Q\approx1.8\pm0.7$.  Given this simple calculation, our observations are consistent with ram pressure stripping playing a prominent role in the removal of gas from star forming galaxies in the ISCS cluster cores.  Currently, similar theoretical predictions do not exist for strangulation, though it too may play a role in SF quenching.  In addition to the simplifying assumptions we've made, we note two caveats: 1) the velocity dispersions of the ISCS clusters are $\sim700$ km s$^{-1}$ \citep{bro11}, lower than the typical velocities at which RPS is thought to be efficient.   As the scatter for the individual galaxy velocities within the ISCS is unknown, the fraction of galaxies for which RPS may be relevant is also unknown.  And 2) hydrodynamical simulations have found that $\sim30$ per cent of a galaxy's hot halo gas may remain intact even 10 Gyr after the initial infall \citep{mcc08}.   

\subsubsection{Mergers and Active Galactic Nuclei}

In Figure~\ref{fig:masslimited1}, we see a striking increase in the average SFR and SSFR of cluster galaxies over the field at high redshift when we examine the cluster outskirts.  Detected at the 5$\sigma$ level, the 250$\mu$m flux in the $\avg{z}=1.4$, $0.5<r<1\,$Mpc bin reveals a $\avg{\textup{SFR}}$ ($\avg{\textup{SSFR}}$) of $\sim3$ ($\sim2$) times the field level at the same redshifts  \citep[though the average SSFR is still within the infrared Main Sequence; ][]{elb11}.     One possible explanation for this enhanced activity is galaxy mergers, which operate in dense environments where galaxy velocities are moderate.  Mergers have been observed at high redshift \citep{bri10, lot11} and a recent study of a z=1.4 cluster using $Herschel$ found that ULIRGs were primarily residing in the cluster outskirts ($r>250\,$kpc), with half of the PACS detected sources showing the disturbed morphologies indicative of merger activity \citep{san13}. 
%An enhanced merger rate has been directly observed in a high redshift ($z=1.62$) cluster \citep{lot12}.

 \citet{man10} presented statistical evidence for rapid mass assembly in the ISCS (consistent with merger activity) by examining the rest-frame 3.6 and 4.5$\mu$m luminosity  functions for cluster galaxies over the redshift range $z=0.3-2$, finding that the characteristic magnitude $m^*$ was well described by passive evolution models up until $z\sim1.4$, above which $m^*$ is abruptly fainter.  This shift in the characteristic 3.6 and 4.5$\mu$m magnitudes, a proxy for the characteristic stellar mass, can be explained by an increase in the merger rate.   These results are corroborated by a study of the SSFR in 16 ISCS clusters between z=1-1.5 using MIPS 24$\mu$m imaging, which finds substantial star formation occurring at all cluster-centric radii and a transition epoch from passively evolving to actively star forming at $z\sim1.4$ \citep{bro13}.  Mergers can both greatly enhance star formation and quickly quench it, as simulations show that mergers often trigger substantial AGN feedback that expels the remaining gas and ends star formation; this process operates on time-scales of $\sim100$ Myr \citep{spr05, hop06, nar10}.  The fraction of AGN has been found to increase by two orders of magnitude within the ISCS sample over z=0-1.5 \citep{gal09, mar13}.   In our sample, we see that the enhanced star formation is occurring primarily in lower mass galaxies,  consistent with the \citet{man10} results and with studies of the merger rate which find that higher mass galaxies ($\gtrsim5\e{10}\Msun$) are undergoing fewer mergers than low mass galaxies \citep{bri10, lot11}.   We note that there may also be minor or dry mergers, even at radii or redshifts where we don't see enhanced star formation activity.

 Though the accretion of galaxy groups on to clusters has also been posited to enhanced star formation and lead to the rapid consumption of gas \citep{mil03, pog04, coi05, fer05}, this process is expected to be more or less continuous over the last 10 billion years \citep{ber09}, which would not explain the abrupt transition from enhanced to quenched that we see in the cluster outskirts (Figure~\ref{fig:masslimited1}). Multiple lines of evidence are pointing toward a prominent role for mergers in the evolution of the ISCS clusters.  Future work using deep $Herschel$ PACS imaging will be used to take a closer look at the radial dependence of the (U)LIRG population in high redshift ISCS clusters (Alberts et al., in preparation). 

%Though multiple lines of evidence are pointing toward a prominent role for mergers in the evolution of the ISCS clusters, further study is needed.  Two studies are in preparation which will help answer the question of mergers in ISCS: targeted deep HST observations will be used to select merger candidates (PI: S. Adam Stanford, xx et al., in prep.) and deep $Herschel$ PACS imaging will be used to take a closer look at the radial dependence of the LIRG/ULIRG populations in high redshift ISCS clusters (Alberts et al., in prep.).

\subsection{Comparison of the Evolution of the SFR in Clusters to Other Studies in the Literature}
\label{sec:compare}

	The most direct comparison to our study is a recent work by \citet{hai13}, who looked at the average SSFRs of massive (M$\gtrsim10^{10} \Msun$) star forming galaxies out to $r_{200}$ in 30 galaxy clusters from $0.15<z<0.3$.  Though their clusters are on average more massive than ours ($\sim10^{14}-10^{15}\Msun$), we probe to similar depths in L$_{\rm IR}$ ($\sim1\e{10}\,\Lsun$) . We find remarkable agreement in that their star forming cluster galaxies also show little differential evolution with respect to the field, but are suppressed below the field level by 28 per cent (see Figure~\ref{fig:redsequence} for comparison).  They further determine that this holds for fixed stellar mass, indicating it is caused by changes in the SFRs at these redshifts.  \citet{hai13} concludes that this systematic reduction of the SFRs in cluster galaxies is due to long timescale ($\gtrsim1\,$Gyr) quenching, such as strangulation or ram pressure stripping.  Combined, our results suggest that the suppression of the SSFRs in star forming cluster galaxies exists over a long redshift baseline ($0.15<z<1.5$), which may indicate a common quenching mechanism in low and high redshift clusters.

	In Section~\ref{sec:redshift}, we quantified the evolution of the average SFR and SSFR as a function of redshift in order to compare to other work in the literature.   We found that, when quantified via the function $y=y_0(1+z)^n$, the average SFR of cluster galaxies goes as $n=5.6\pm0.6$ in the cluster cores and $n=5.9\pm1.0$ in the core+outskirts.   We compare the evolution of the average SFR to two popular quantities in the literature: the total SFR per halo mass, $\Sigma({\rm SFR})$/M$_{\textup halo}$, which is particularly useful measurement for comparing systems of different mass, and the fraction of star-forming galaxies, $f_{SF}$.  Several studies have found that the redshift dependence of the total SFR per halo mass goes as $n\sim5-7$ \citep{kod04, fin04, fin05, gea06, bai09, chu10, koy10, hay11, pop12, web13, hai13}.  Given that our cluster sample is uniform in mass across our redshift range, we can fairly compare the evolution of our average SFR to this quantity.  The evolution of $f_{SF}$ is somewhat less constrained with $f_{SF}\propto(1+z)^{2-7}$ \citep{kod04, gea06, sai08, bai09, hai09, web13, hai13}.  Comparing to this quantity is interesting, however, given the suggestion that the evolution we see in the SF activity in our full cluster galaxy population is dominated by the changing fraction of star forming galaxies. Comparisons between our results and these literature results are complicated as we are measuring different quantities and have different cluster masses, cluster selection, galaxy selections, SFR tracers, and redshift ranges.    Nevertheless, we find good agreement between our measured evolution of the average SFR and the measured evolution of both $\Sigma({\rm SFR})$/M$_{\textup halo}$ and $f_{SF}$ from previous works.   In particular, we note the high redshift cluster studies of \citet{web13}, who looked at IR-luminous (L$_{\rm IR}>2\e{11}\Lsun$) galaxies in 42 red-sequence selected cluster from $0.3<z<1$.  They found evolutions of $n=5.4\pm1.5$ for the total SFR per halo mass and $n=5.1\pm1.9$ for the star forming fraction.  Given this consistent evolution between these quantities and between our studies (and others), this indicates that the total SFR per halo mass in cluster galaxies could be tightly correlated with the star forming fraction in clusters over a range of luminosities and that different cluster samples may be experiencing similar quenching mechanisms over a range of redshifts.

At $z>1$, our findings provide important direct support for conclusions drawn from previous investigations of the ISCS clusters.  In particular, we have shown field-level star formation rates, indicating ongoing stellar mass assembly, at $z>1.2$, matching the inference based on the near-IR luminosity function evolution of cluster members by \citet{man10}.  In addition, we have shown that, at $z\lesssim1.2$, one or more processes are rapidly halting star formation in some of these cluster galaxies (Figure~\ref{fig:masslimited05}-\ref{fig:masslimited1}).  In combination, these scenarios can explain the nearly constant colour of the optically defined quiescent galaxies in ISCS clusters \citep{sny12}:  at any given time, the population of red cluster galaxies reflects the extended star formation histories of the previous star forming galaxies that have been very rapidly quenched in their past, possibly in a stochastic manner.  Therefore we conclude that there is broad agreement between the scenarios implied by the stellar mass build-up of cluster galaxies, the apparent stellar age evolution of cluster ellipticals, and the SFRs of cluster galaxies as measured directly \citep[this work;][]{bro13}.

\section{Conclusions}
\label{sec:conclusions}

In this work, we have used a large, uniform cluster sample over a long redshift baseline (z=0.3-1.5) in order to analyse the star formation activity in cluster galaxies relative to the field as a function of cosmic time.  Through a stacking analysis, we have probed to low infrared luminosities and determined the average L$_{\rm IR}$, SFRs, and SSFRs by measuring the average 250$\mu$m flux of mass-limited samples of thousands of cluster galaxies and tens of thousands of field galaxies.  Using robust, statistical methods, we have accounted for source blending/clustering bias and field galaxy contamination (due to photometric redshift uncertainties) in our cluster galaxy stacking.  Our main results are as follows.

\begin{enumerate}
\item Our full (star-forming and quiescent) cluster galaxy sample exhibits rapid evolution with cosmic time as compared to the field.  We quantify this evolution as an exponential function of time and find that cluster galaxies in the cluster cores  ($r<0.5\,$Mpc) have an e-folding time of $\sim1.5\,$Gyr, as compared to $\sim2.4\,$Gyr for field galaxies.  The average SFR in the cluster cores is quenched below the field level for much of cosmic time ($\sim9$ billion years) but draws even with the field at $z>1.2$.  When accounting for stellar mass by measuring the SSFR, the core cluster galaxies don't quite drawn even with the field up to $z\sim1.5$, but still show a statistically faster evolution than the average SSFR of field galaxies (see Table~\ref{tbl:fit}).  In the cluster outskirts ($0.5<r<1\,$Mpc), we see enhanced SFRs (SSFRs) of $\sim3$ ($\sim2$) times the field level at $\avg{z}$=1.4, likely due to increased merger activity among the infalling galaxy population.  These results confirm the transition epoch toward active star formation and mass assembly at $z\sim1.4$ seen in previous studies.

\item  When divided into lower and higher mass bins, we see that the SSFRs of the higher mass galaxies (M$_{\star}>6.3\e{10}\Msun$) in the cluster cores are quenched below the field level, but otherwise show no strong differential evolution relative to the field.  We suggest that strangulation from the hot cluster ICM is responsible for the lower level of star formation, but that the overall evolution with time of the higher mass cluster galaxies is dominated by the same mechanism as higher mass field galaxies, i.e. mass-quenching.  Lower mass galaxies ($1.3\e{10}<$M$_{\star}<6.3\e{10}\Msun$) seem to be driving the differing evolution from the field galaxies in the cluster cores with SSFRs that begin reaching the field level at $z>1.2$.  In the outskirts, both mass bins show a more rapid evolution in the clusters than the field, with lower mass galaxies showing enhanced  SF at $\avg{z}$=1.4, which may suggest that lower mass galaxies are preferentially experiencing major mergers which trigger starbursts.  %The disparate rates of evolution for the two mass bins in both the cluster cores and outskirts suggest that downsizing may be occurring on cluster scales as well as in the field.

\item We find that though the $\avg{\textup{SFR}}$ of blue, star forming galaxies decreases faster than blue galaxies in the field, the SSFR of blue galaxies shows the same behaviour as high mass galaxies (suppressed below the field but with no strong differential evolution). The exception is the the cluster outskirts at  $\avg{z}$=1.4, where blue galaxies show enhanced SF activity.  This suggests that environment could be strongly effecting the SFRs and/or stellar mass distributions of blue, star forming galaxies in clusters.

\item We suggest that our results are consistent with both strangulation and ram pressure stripping operating in these clusters, and increased merger activity occurring in the cluster outskirts at high redshift.  Strangulation, a long timescale process, may particularly be affecting high mass galaxies in the cluster cores.  Ram pressure stripping, a shorter timescale process, may control our fraction of star forming to quiescent galaxies, driving the trend that we see in the full cluster sample.  Mergers and AGN provide a natural explanation for enhanced star formation activity and quenching on short time-scales in the cluster outskirts at $\avg{z}$=1.4.  

%\item In stacking field galaxies as a function of redshift, we found a break in the evolution of the SFRs and SSFRs at $z\sim0.8$, with lower redshift field galaxies showing a more rapid quenching than at high redshift.  This is reminiscent of the differential ramp up of LIRGs and ULIRGs with time and the general form of the star formation rate density of the Universe \citep[see][]{mag13}.  We suggest that this break might be related to the changing efficiency of cold mode accretion over time, which we will explore further in a future work.
\end{enumerate}

This study has probed the average star formation properties of cluster galaxies relative to the field using a large cluster sample over a wide range in redshift.  Individual cluster galaxy SF properties will be examined for high redshift ($z$=1-2) ISCS clusters using deep $Herschel$ PACS (PI: Alexandra Pope) imaging in future work (Alberts et al., in preparation).

\section*{Acknowledgements}
We would like to thank our colleagues in the IRAC Shallow Survey, IRAC Shallow Cluster Survey, NDWFS, SDWFS, and MAGES teams, as well as the HerMES collaboration for making their data publicly available.   In addition, the authors extend a special thanks to Marco Viero for his contributions.  The authors thank the anonymous referee for his/her helpful comments and suggestions.  AP and AD acknowledge the hospitality of the Aspen Center for Physics, which is supported by the National Science Foundation Grant No. PHY-1066293. AD's research activities are supported by NOAO, which is operated by the Association of Universities for Research in Astronomy (AURA) under cooperative agreement with the NSF.  This work is based on observations made with $Herschel$, a European Space Agency Cornerstone Mission with significant participation by NASA. Support for this work was provided by NASA through an award issued by JPL/Caltech.  We additionally thank the $Herschel$ Helpdesk for their assistance in the data reduction process.  SPIRE has been developed by a consortium of institutes led by Cardiff University (UK) and including Univ. Lethbridge (Canada); NAOC (China); CEA, LAM (France); IFSI, Univ. Padua (Italy); IAC (Spain); Stockholm Observatory (Sweden); Imperial College London, RAL, UCL-MSSL, UKATC, Univ. Sussex (UK); and Caltech, JPL, NHSC, Univ. Colorado (USA). This development has been supported by national funding agencies: CSA (Canada); NAOC (China); CEA, CNES, CNRS (France); ASI (Italy); MCINN (Spain); SNSB (Sweden); STFC (UK); and NASA (USA).  This work is additionally based on observations made with the Spitzer Space Telescope, which is operated by the Jet Propulsion Laboratory, California Institute of Technology under a contract with NASA. Support for this work was provided by NASA through an award issued by JPL/Caltech.

\label{bib}

\appendix
\section{Description of Bo\"{o}tes SPIRE Maps}
\label{appendix:a}

Here we describe our reduction of the $Herschel$ SPIRE observations of the Bo\"{o}tes field, publicly available from HerMES \citep{oli12}.  The Bo\"{o}tes field was observed with 9 AORs in SPIRE PACS Parallel Mode between December 3, 2009 and January 1, 2010.  A listing of the observation IDs can be seen in Table~\ref{tbl:obsids}.  Five of the Astronomical Observation Requests (AORs) cover the central 2 square degrees of the field, while the other four AORS cover half of the full $\sim$8 square degrees centered on 14:32:06 +34:16:48.  At least two AORs overlap in each area of the map.  Very few cluster galaxies are detected in the SPIRE maps; because of this, this work has focused on stacking analyses in order to probe the average SF in all cluster galaxies.  We describe here additional details about the source catalogs and associated simulations in order to validate the map and flux measurements used in this study. 

\begin{table}
\begin{minipage}[!ht]{\textwidth}
\caption{Summary of Bo\"{o}tes AORs.}
\label{tbl:obsids}
\begin{tabular}{lcccc}
\hline
ObsID & Exposure Time & Operation & RA & Dec  \\
& [seconds] & Day (OD) & &  \\
\hline
1342187711 & 14809.0 &  203 & 14:32:07 & +34:16:55 \\
1342187712 & 14809.0 & 203 & 14:32:11 & +34:17:36 \\
1342187713 &  14809.0 & 203 & 14:32:09  & +34:17:28  \\
1342188090 & 14809.0 & 214 & 14:32:05 & +34:16:36  \\
1342189108 & 14374.0 & 240 & 14:32:02  & +34:19:14 \\
1342188650 &  24748.0 & 228 & 14:35:30 & +34:25:35 \\
1342188651 & 24748.0 & 228 & 14:28:42  & +34:09:38  \\
1342188681 &  24748.0 & 229 & 14:32:17 & +33:33:44  \\
1342188682 & 24748.0 & 229 & 14:31:35  & +34:59:08  \\
\hline
\end{tabular}
\\
%\tiny{$^a$Not corrected for field contamination (see Section~\ref{sec:clusterstacks}).}\\
%\tiny{$^b$AGN classification is determined based on X-ray and IRAC \\
%colors \citep[][]{kir13}. }
\end{minipage}
\end{table}

\subsection{Data Reduction and Catalogs}

Data reduction was done using \lowercase{HIPE} version 7 \citep{ott10}.  The 9 AORs were reduced separately up to Level 2 following the standard pipeline with two exceptions:  (1) deglitching was performed using the more advanced sigma-kappa deglitcher rather than the default wavelet deglitcher.  The sigma-kappa deglitcher uses an iterative process to reject outliers after adaptive highpass filtering of the signal timeline.  The final error maps were examined for bright pixels which may indicate missed glitches and the glitch mask was adjusted accordingly.  (2)  Due to striping in the maps, a high order polynomial baseline removal was used instead of the default median baseline removal.  Final, level 2 maps of each AOR were produced using the naive mapmaker.  The calibration tree used was {\lowercase{\sc spire\textunderscore cal\textunderscore 7\textunderscore 0}}.

Astrometry corrections were derived from stacking bright MAGES 24$\mu$m sources on the individual, reduced AORs.   The stacked images were fit with the SPIRE PSF, from which typical offsets of $\sim1$ arcseconds in RA and $\sim2$ arcseconds in Dec were determined.  Mosaicking was performed on the Level 1 scans (which include deglitching) of all 9 AORs for each waveband (250, 350, and 500$\mu$m) after the application of astrometry corrections and polynomial baseline removal.  The error maps generated with each mosaic are the standard deviation of the data points falling into a given pixel and represent the associated instrument noise.  Using the error maps, the 5$\sigma$ depths of the inner (outer) portion of the 250, 350, and 500$\mu$m maps are 14.5, 11.5, and 14.5 mJy (26.5, 21.5, and 26.0 mJy).  This does not include confusion noise, which is 5.8, 6.2, and 6.8 mJy beam$^{-1}$ for 250, 350, and 500$\mu$m \citep{ngy10}.  These values are summarized in Table~\ref{tbl:depths}.

The maps were post-processed using a matched-filter technique which optimizes the S/N ratio for confusion-dominated submillimetre maps and improves source deblending by convolving the maps with a Gaussian which is narrower than the PSF of the observations \citep[see][]{cha11}.  Source finding was performed by identifying local maxima in both the unfiltered (UF) and matched-filtered (MF) maps.  SPIRE are normalized to have a zero mean baseline and units of Jy beam$^{-1}$.  This means that the flux density of the peak pixel provides an accurate estimation of the integrated flux density of a source.  The instrument noise associated with each source is given by the corresponding pixel in the error map.  The source detection threshold was set at 5$\sigma$, as determined by the error maps (which do not take into account confusion noise).  To determine sub-pixel source locations, each detected source was weighed by the S/N in the surrounding pixels.  

Extended sources were identified by eye and masked out if at least one axis exceeded 1.5 times the FWHM of the SPIRE beam.  Thirteen extended sources were identified.  In addition, a 4.8 square arcminute rectangular area centered on 14:33:11.8, +33:26:27 was masked in all maps due to bad pixels in the 500$\mu$m map.  Point source catalogs were generated after masking out the extended sources and bad pixel region.   The $5\sigma$ catalogs for the unfiltered maps contain 14,356, 10,641, and 3,437 point sources for 250, 350, and 500$\mu$m.  The 5$\sigma$ catalogs for the matched-filter maps have 21,892, 13,692, and 5,137 point sources.

\begin{table*}
\begin{minipage}[!ht]{\linewidth}
\caption{SPIRE Map Statistics}
\label{tbl:depths}
\begin{tabular}{lccccccccc}
\hline
Wavelength & FWHM & Pixel Size & \multicolumn{2}{c}{5$\sigma$ Depth [mJy]} & Confusion &\multicolumn{2}{c}{90$\%$ Comp. (UF) [mJy]} & \multicolumn{2}{c}{90$\%$ Comp. (MF) [mJy]} \\
& [arcsec] & [arcsec] pixel$^{-1}$ & Inner & Outer & Noise$^{a}$ (1$\sigma$) [mJy] & Inner & Outer & Inner & Outer  \\
\hline
250$\mu$m & 18 & 6 & 14.5 & 26.5 & 5.8 & 20 & 33 & 18 & 25  \\
350$\mu$m & 25 & 10 & 11.5 & 21.5 & 6.2 & 20 & 35 & 18 & 28  \\
500$\mu$m & 36 & 14 & 14.5 & 26.0 & 6.8 & 22 & 35 & 22 & 28 \\
\hline
\end{tabular}
\\
\tiny{$^a$\citet{ngy10}}\\
\end{minipage}
\end{table*}

\subsection{Completeness Testing}

Completeness simulations were performed on all six maps (UF and MF) for 250, 350, and 500$\mu$m in order to quantify completeness, positional uncertainties, and flux boosting.  As the $Herschel$ SPIRE PSF is nearly Gaussian, fake sources were inserted directly into the maps as Gaussians with the appropriate FWHM and with a peak value scaled to the desired flux.  Inserting fake sources directly into the real map accounts for all sources of noise, including confusion noise.  Given the size of the Bo\"{o}tes field, 100 sources can be inserted into both the inner and outer regions at a time without significantly altering the properties of the original map.  We impose the restriction that no two fake sources can be placed within 100 arcseconds of each other.  Fake sources were given fluxes ranging from 6-10 mJy in steps of 2 mJy, 10-80 mJy in steps of 5 mJy and 100-200 mJy in steps of 100 mJy and placed in random positions.  We generated 100 simulated maps, each with 100 fake sources in both the inner and outer regions, per flux bin per wavelength for the UF and MF maps.

In order to determine the recovery search radius, 10,000 random apertures of increasing size were placed on the UF and MF maps to find the radius at which there is a 5$\%$ chance of randomly encountering a detected source in the inner region (which is more crowded due to its depth).  The search radius adopted from this is 10$\arcsec$, 12$\arcsec$, and 18$\arcsec$ for 250, 350, and 500$\mu$m for both the UF and MF maps.  In the simulated maps, fake sources are then searched for using the appropriate search radius at their original location.  A source is considered to be recovered if it is detected at $\geq5\sigma$ and its  position and flux are recorded as a function of input flux.

\begin{figure}
\centering
\includegraphics[width=1\linewidth, trim=0 0 0 0, clip]{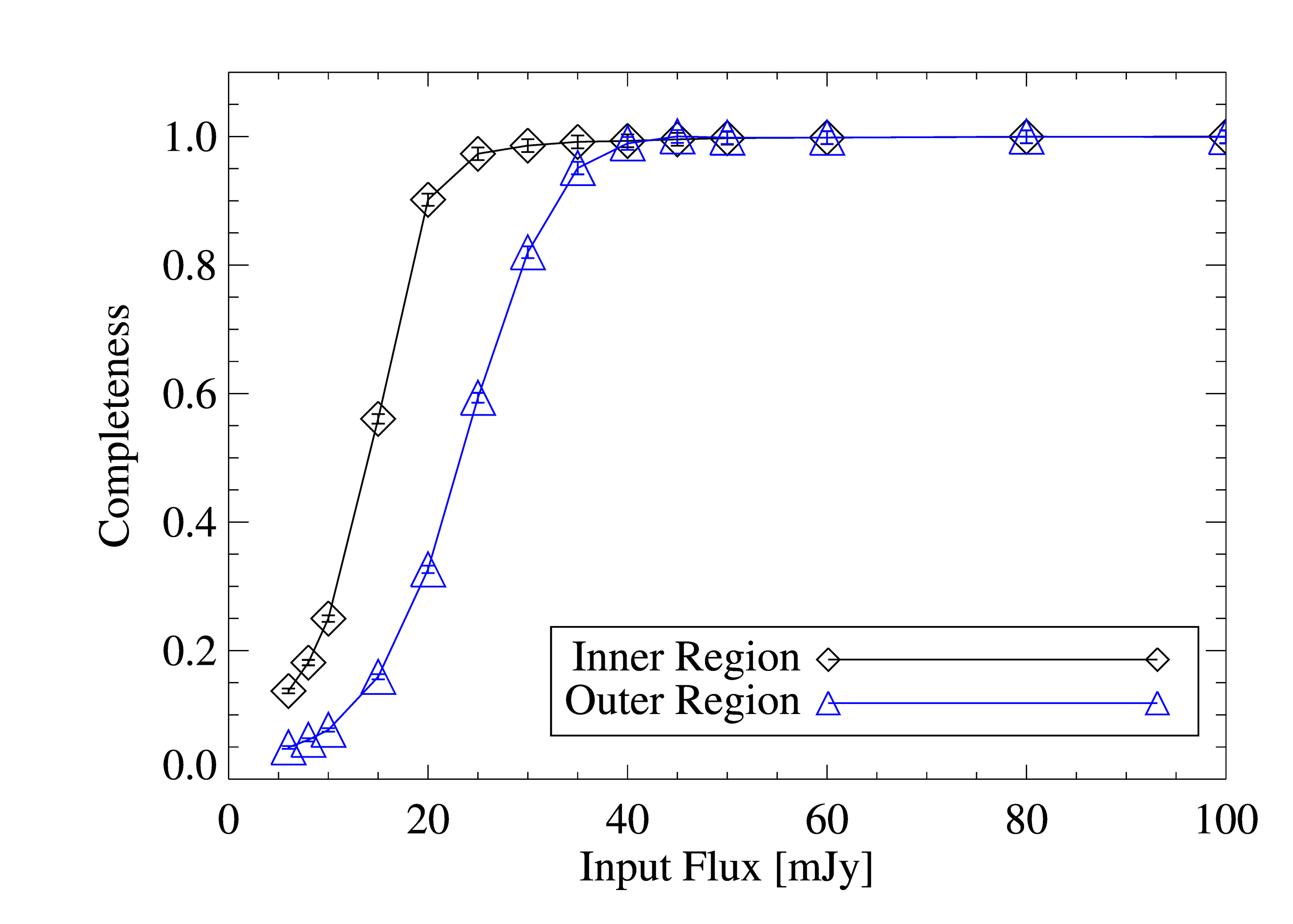}
\caption{The completeness as a function of flux for fake sources inserted into the Bo\"{o}tes 250$\mu$m unfiltered map.  The inner region (black diamonds) is $\sim$2 times deeper than the outer region (blue triangles).  The errors are Poisson errors.}
\label{fig:comp}
\end{figure}

\begin{figure}
\centering
\includegraphics[width=1\linewidth, trim=0 0 0 0, clip]{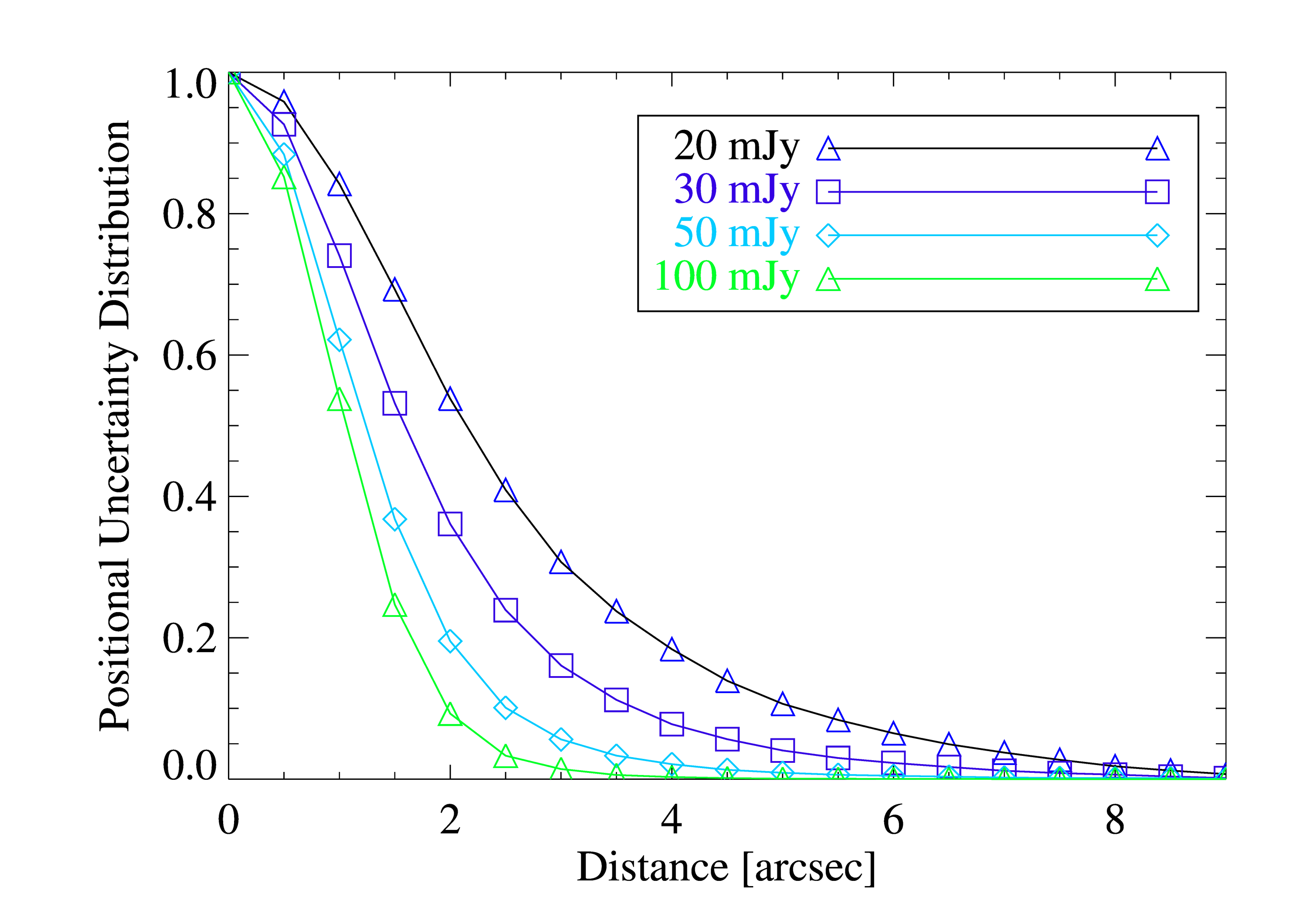}
\caption{The positional uncertainty distribution as a function of distance $D$ for the inner region of the Bo\"{o}tes 250$\mu$m unfiltered map.  This function indicates the probability that a source will be detected at distance greater than $D$ from it's true position as a function of source flux.}
\label{fig:pu}
\end{figure}

The 90$\%$ completeness fluxes are listed in Table~\ref{tbl:depths} and an example of the completeness as a function of input flux for the 250$\mu$m unfiltered map can be seen in Figure~\ref{fig:comp}.  For fake sources which are recovered, we calculate the distance between the position at which the source is recovered and its original position and determine the probability, $P(>D;S)$, that a SPIRE source will be detected at a distance greater than $D$ from its true position as a function of the source's flux.  The positional uncertainties for several source fluxes can be seen for the inner region of the 250$\mu$m unfiltered map in Figure~\ref{fig:pu}.  For a 20 mJy source, the probability that it will be detected within 7\arcsec, 8.5\arcsec, and 13\arcsec for the 250, 350, and 500$\mu$m for the UF maps and 5.5\arcsec, 8\arcsec, and 12\arcsec for the MF maps is $\geq90\%$.  In addition to the positional uncertainties, we quantified flux boosting across the map due to source blending.  The recovered fluxes of the fake sources were compared to their input flux as a function of input flux.   We found that flux boosting is negligible for sources $\geq20\,$mJy for all maps and rises steeply with decreasing flux.

\label{lastpage}

\end{document}